\documentclass[conference,letterpaper]{IEEEtran}

\usepackage[utf8]{inputenc} 
\usepackage[T1]{fontenc}
\usepackage{url}
\usepackage{ifthen}
\usepackage{cite}
\usepackage[cmex10]{amsmath} 

\usepackage{amssymb,amsfonts}

\usepackage{lipsum}
\usepackage{mathtools}
\usepackage{cuted}

\usepackage{balance}

\usepackage{amsmath}
\DeclareMathOperator{\tr}{tr}

\newtheorem{theorem}{\it Theorem}

\newtheorem{definition}{\it Definition}

\newtheorem{corollary}{\it Corollary}
\newtheorem{proposition}{\it Proposition}
\newtheorem{example}{\it Example}

\begin{document}
\title{Feedback Capacity of Parallel ACGN Channels and Kalman Filter: Power Allocation with Feedback}


\author{%
  \IEEEauthorblockN{Song Fang and Quanyan Zhu}
  \IEEEauthorblockA{Department of Electrical and Computer Engineering, New York University, New York, USA\\
  	Email: \{song.fang, quanyan.zhu\}@nyu.edu}
}


\maketitle


\begin{abstract}
	
	In this paper, we relate the feedback capacity of parallel additive colored Gaussian noise (ACGN) channels to a variant of the Kalman filter. By doing so, we obtain lower bounds on the feedback capacity of such channels, as well as the corresponding feedback (recursive) coding schemes, which are essentially power allocation policies with feedback, to achieve the bounds. The results are seen to reduce to existing lower bounds in the case of a single ACGN feedback channel, whereas when it comes to parallel additive white Gaussian noise (AWGN) channels with feedback, the recursive coding scheme reduces to a feedback ``water-filling'' power allocation policy.

\end{abstract}


\section{Introduction}

The feedback capacity \cite{cover1989gaussian} of additive colored Gaussian noise (ACGN) channels has been a long-standing problem in information theory, generating numerous research papers over the years, due to its significance in understanding and applying communication/coding with feedback.
In general, we refer to the breakthrough paper \cite{kim2010feedback} and the references therein for a rather complete literature review; see also \cite{kim2006feedback, kim2006gaussian} for possibly complementary paper surveys. Meanwhile, papers on this topic have also been coming out continuously after \cite{kim2010feedback}, which include but are certainly not restricted to \cite{ardestanizadeh2012control, liu2014convergence, liu2015capacity, stavrou2017sequential,
	liu2018feedback,
	li2018youla, rawat2018computation, pedram2018some,
	kourtellaris2018information, gattami2018feedback, ihara2019feedback, fang2020connection, aharoni2020capacity}.
Most of the aforementioned works, however, focused merely on the feedback capacity of a single ACGN channel, so to speak, whereas when it comes to parallel ACGN channels, the corresponding results have been lacking in general. One exception is the recent paper \cite{rawat2018computation} which generalized the computational approach in \cite{li2018youla} to multi-antenna ACGN channels.

In this paper, we establish a connection between a parallel of ACGN channels with feedback and a variant of the multi-input multi-output (MIMO) Kalman filter for colored Gaussian noises. In light of this, we obtain lower bounds on feedback capacity for parallel ACGN channels by examining the algebraic Riccati equations associated  with the Kalman filter. Meanwhile, the Kalman filtering systems, which are essentially feedback (closed-loop) systems, naturally provide recursive coding schemes, in terms of feedback power allocation policies, to achieve the lower bounds. In addition, the lower bounds are shown to be consistent with existing feedback capacity results when it comes to a single ACGN channel. It is also seen that in the special case of parallel additive white Gaussian noise (AWGN) channels, the recursive coding reduces to a feedback ``water-filling'' solution.

In a broad sense, in this paper we utilize a control-theoretic approach towards this problem as in, e.g., \cite{Eli:04, yang2007feedback,  tatikonda2008capacity, ardestanizadeh2012control, liu2014convergence, liu2015capacity,
	li2018youla, rawat2018computation,
	pedram2018some, kourtellaris2018information, gattami2018feedback,  fang2020connection}; see also \cite{fang2017towards} and the references therein.
Note also that although the organization of this paper resembles that of \cite{fang2020connection} to a certain extent, the results  are not trivial generalizations of those therein, as evidenced by the results themselves as well as their proofs. 


The rest of the paper is organized as follows. Section~II provides the preliminary background on feedback capacity and Kalman filter. In Section~III, we present the main results of this paper. Concluding remarks are given in Section~IV.

\section{Preliminaries}

In this paper, we consider real-valued continuous random variables and discrete-time stochastic processes they compose. All random variables and stochastic processes are assumed to be zero-mean for simplicity and without loss of generality. We represent random variables using boldface letters. The logarithm is defined with base $2$.
A stochastic process $\left\{ \mathbf{x}_{k}\right\}$ is said to be asymptotically stationary if it is stationary as $k \to \infty$, and herein stationarity means strict stationarity \cite{Pap:02}. Note in particular that, for simplicity and
with abuse of notations, we utilize $\mathbf{x} \in \mathbb{R}$ and $\mathbf{x} \in \mathbb{R}^n$ to
indicate that $\mathbf{x}$ is a real-valued random variable and that $\mathbf{x}$
is a real-valued $n$-dimensional random vector, respectively.

The following definitions of entropy and entropy rate are adapted from, e.g., \cite{Cov:06}. 

\begin{definition} The differential entropy of a random vector $\mathbf{x}$ with density $p_{\mathbf{x}} \left(x\right)$ is defined as
	\begin{flalign}
	h\left( \mathbf{x} \right)
	=-\int p_{\mathbf{x}} \left(x\right) \log p_{\mathbf{x}} \left(x\right) \mathrm{d} x. \nonumber
	\end{flalign}
	The entropy rate of a stochastic process $\left\{ \mathbf{x}_{k}\right\}$ is defined as
	\begin{flalign}
	h_\infty \left(\mathbf{x}\right)=\limsup_{k\to \infty} \frac{h\left(\mathbf{x}_{0,\ldots,k}\right)}{k+1}. \nonumber
	\end{flalign}
\end{definition}

\subsection{Feedback Capacity} \label{notation}

Consider a parallel of $n$ additive colored Gaussian noise channels given by
\begin{flalign}\mathbf{y}_{k} = \mathbf{x}_{k} + \mathbf{z}_{k}, \nonumber
\end{flalign}
where $\left\{ \mathbf{x}_{k} \right\}, \mathbf{x}_{k} \in \mathbb{R}^n$ denotes the channel input, $\left\{ \mathbf{y}_{k} \right\}, \mathbf{y}_{k} \in \mathbb{R}^n$ denotes the channel output, and $\left\{ \mathbf{z}_{k} \right\}, \mathbf{z}_{k} \in \mathbb{R}^n$ denotes the additive noise which is assumed to be stationary colored Gaussian. The feedback capacity $C_{\text{f}}$ of such a channel with power constraint $\overline{P}$ is given by \cite{cover1989gaussian, Cov:06}
\begin{flalign} \label{fcdef}
C_{\text{f}} = \sup_{ \lim_{k\to \infty} \frac{1}{k+1} \sum_{i=0}^{k} \tr \mathbb{E} \left[  \mathbf{x}_{i}  \mathbf{x}^{\mathrm{T}}_{i} \right] \leq \overline{P}} \left[ h_{\infty} \left( \mathbf{y} \right) -  h_{\infty} \left(  \mathbf{z} \right) \right].
\end{flalign}

It is worth mentioning that in the case of $n=1$, it was shown in \cite{kim2010feedback} that the optimal channel input process $\left\{ \mathbf{x}_{k} \right\}$ is stationary and in the form of
\begin{flalign} 
\mathbf{x}_{k} = \sum_{i=1}^{\infty} b_{i} \mathbf{z}_{k-i},~b_{i} \in \mathbb{R}, \nonumber
\end{flalign} 
while satisfying $\mathbb{E} \left[  \mathbf{x}_{k}^2 \right] \leq \overline{P}$. This has also been generalized to the case of $n$ parallel channels by \cite{rawat2018computation}. 
For the purpose of this paper, however, it suffices to consider the original definition in \eqref{fcdef}.

\subsection{Kalman Filter} \label{kalamnsection}


\begin{figure}
	\begin{center}
		\vspace{-3mm}
		\includegraphics [width=0.5\textwidth]{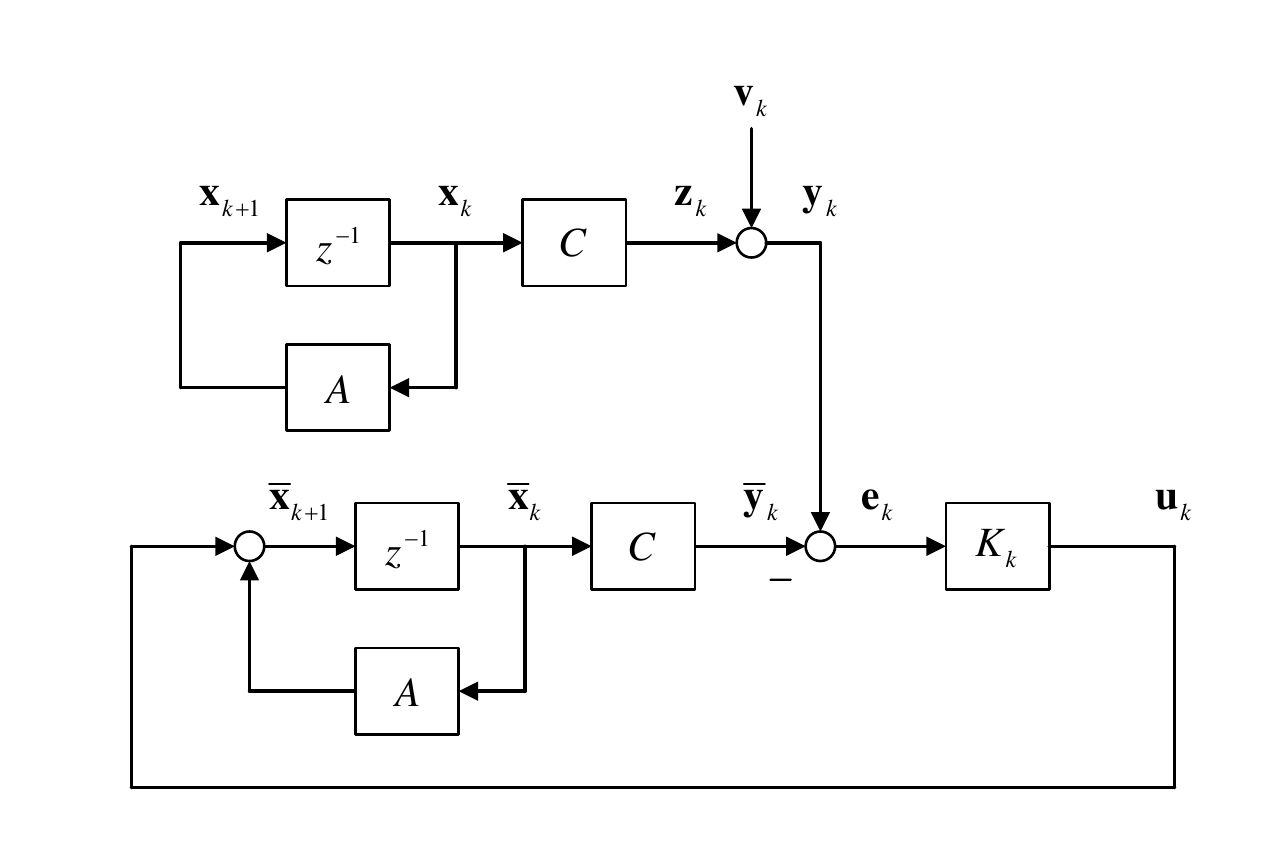}
		\vspace{-6mm}
		\caption{The Kalman filtering system.}
		\label{kalman1}
	\end{center}
	\vspace{-3mm}
\end{figure}

We now give a brief review of (a special case of) the MIMO Kalman filter \cite{linearestimation, anderson2012optimal}; note that hereinafter the notations are not to
be confused with those in Section~\ref{notation}. Particularly, consider the Kalman filtering system depicted in Fig.~\ref{kalman1}, where the state-space model of the plant to be estimated is given by
\begin{flalign} \label{plant}
\left\{ \begin{array}{rcl}
\mathbf{x}_{k+1} & = & A \mathbf{x}_{k},\\
\mathbf{y}_{k} & = & C \mathbf{x}_{k} + \mathbf{v}_k.
\end{array} \right. 
\end{flalign}
Herein, $\mathbf{x}_{k} \in \mathbb{R}^n$ is the state to be estimated, $\mathbf{y}_{k} \in \mathbb{R}^n$ is the plant output, and $\mathbf{v}_{k} \in \mathbb{R}^n$ is the measurement noise, whereas the process noise, normally denoted as $\left\{ \mathbf{w}_{k} \right\}$ \cite{linearestimation, anderson2012optimal}, is assumed to be absent. The system matrix is $ A \in \mathbb{R}^{n \times n}$ while the output matrix is $ C \in \mathbb{R}^{n \times n}$, and we assume that $A$ is anti-stable (i.e., all the eigenvalues are unstable with magnitudes greater than or equal to $1$) while the pair $\left( A, C \right)$ is observable (and thus detectable \cite{astrom2010feedback}). Suppose that $\left\{ \mathbf{v}_{k} \right\}$ is white Gaussian with covariance $V = \mathbb{E} \left[ \mathbf{v}_{k} \mathbf{v}_{k}^{\mathrm{T}} \right] \succ 0$ and the initial state $\mathbf{x}_{0}$ is Gaussian with covariance $ \mathbb{E} \left[ \mathbf{x}_0 \mathbf{x}_0^{\mathrm{T}} \right] \succ 0$. Furthermore, $\left\{ \mathbf{v}_{k} \right\}$ and $\mathbf{x}_{0}$ are assumed to be uncorrelated.  
Correspondingly, the Kalman filter (in the observer form \cite{astrom2010feedback}) for \eqref{plant} is given by
\begin{flalign} \label{estimator}
\left\{ \begin{array}{rcl}
\overline{\mathbf{x}}_{k+1}&=&A \overline{\mathbf{x}}_{k} +\mathbf{u}_k, \\
\overline{\mathbf{y}}_{k}&= & C \overline{\mathbf{x}}_{k}, \\
\mathbf{e}_{k}&=&\mathbf{y}_{k}-\overline{\mathbf{y}}_k, \\
\mathbf{u}_{k}&=& K_{k} \mathbf{e}_{k},
\end{array} 
\right.
\end{flalign}
where $\overline{\mathbf{x}}_{k} \in \mathbb{R}^n$, $\overline{\mathbf{y}}_{k} \in \mathbb{R}^n$, $\mathbf{e}_{k} \in \mathbb{R}^n$, and $\mathbf{\mathbf{u}}_{k} \in \mathbb{R}^n$. Herein, $K_{k}$ denotes the observer gain \cite{astrom2010feedback} (note that the observer gain is different from the Kalman gain by a factor of $A$; see, e.g.,  \cite{anderson2012optimal, astrom2010feedback} for more details) given by
\begin{flalign}
K_{k} = A P_{k} C^{\mathrm{T}} \left( C P_{k} C^{\mathrm{T}} + V \right)^{-1}, \nonumber
\end{flalign}
where $P_k$ denotes the state estimation error covariance as
\begin{flalign}
P_k = \mathbb{E} \left[ \left(\mathbf{x}_k -\overline{\mathbf{x}}_k \right) \left(\mathbf{x}_k -\overline{\mathbf{x}}_k \right)^{\mathrm{T}} \right]. \nonumber
\end{flalign}
In addition, $P_k$ can be obtained iteratively by the Riccati equation
\begin{flalign}
P_{k+1}
= A P_{k} A^{\mathrm{T}} - A P_{k} C^{\mathrm{T}} \left( C P_{k} C^{\mathrm{T}} + V \right)^{-1} C P_{k} A^{\mathrm{T}} \nonumber
\end{flalign}
with $P_{0} = \mathbb{E} \left[ \mathbf{x}_0 \mathbf{x}_0^{\mathrm{T}} \right] \succ 0$.
Additionally, it is known \cite{linearestimation, anderson2012optimal} that when $\left( A, C \right)$ is detectable, the Kalman filtering system converges, i.e., the state estimation error $\left\{ \mathbf{x}_k -\overline{\mathbf{x}}_k \right\}$ is asymptotically stationary. Moreover, in steady state, the optimal state estimation error variance
\begin{flalign}
P =\lim_{k\to \infty} \mathbb{E} \left[ \left(\mathbf{x}_k -\overline{\mathbf{x}}_k \right) \left(\mathbf{x}_k -\overline{\mathbf{x}}_k \right)^{\mathrm{T}} \right] \nonumber
\end{flalign} 
attained by the Kalman filter is given by the (non-zero) positive semi-definite solution \cite{anderson2012optimal} to the algebraic Riccati equation
\begin{flalign} \label{are}
P
= A P A^{\mathrm{T}} - A P C^{\mathrm{T}} \left( C P C^{\mathrm{T}} + V \right)^{-1} C P A^{\mathrm{T}}, 
\end{flalign} 
whereas the steady-state observer gain is given by 
\begin{flalign}  \label{gain}
K = A P C^{\mathrm{T}} \left( C P C^{\mathrm{T}} + V \right)^{-1}.
\end{flalign}

\begin{figure}
	\begin{center}
		\vspace{-3mm}
		\includegraphics [width=0.5\textwidth]{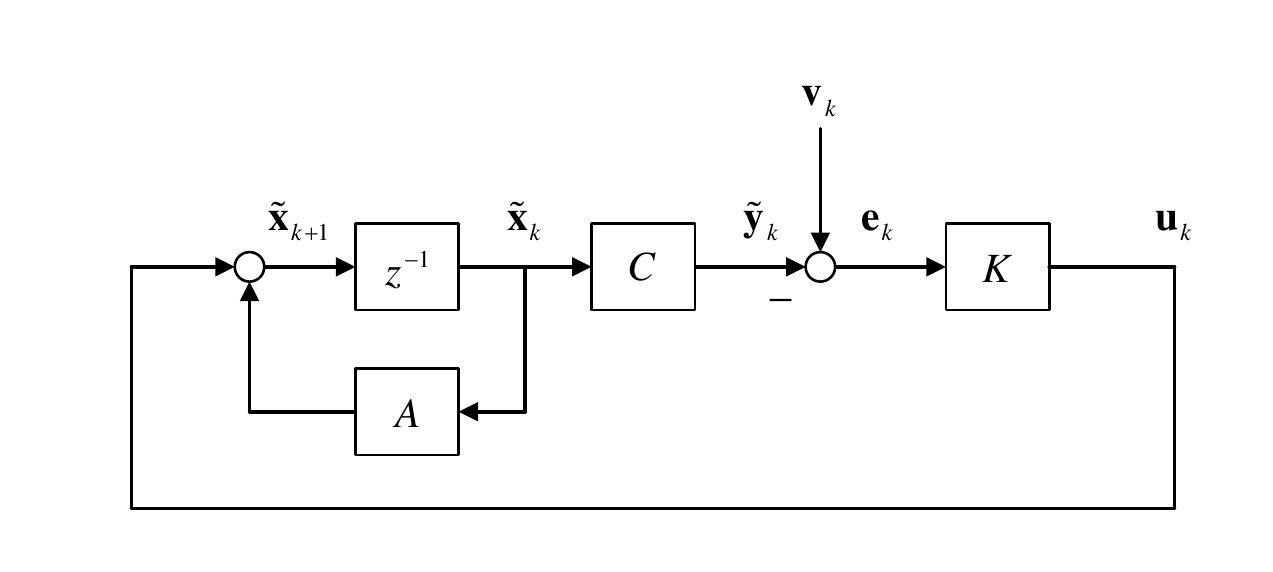}
		\vspace{-6mm}
		\caption{The steady-state Kalman filtering system in integrated form.}
		\label{kalman2}
	\end{center}
	\vspace{-3mm}
\end{figure}

In fact, by letting $\widetilde{\mathbf{x}}_{k} = \overline{\mathbf{x}}_{k} - \mathbf{x}_{k} $ and $\widetilde{\mathbf{y}}_{k} = \overline{\mathbf{y}}_{k} - \mathbf{z}_{k} = \overline{\mathbf{y}}_{k} - C \mathbf{x}_{k} $, we may integrate the steady-state systems of \eqref{plant} and \eqref{estimator} into an equivalent form:
\begin{flalign}
\left\{ \begin{array}{rcl}
\widetilde{\mathbf{x}}_{k+1}&=&A \widetilde{\mathbf{x}}_{k} +\mathbf{u}_k, \\
\widetilde{\mathbf{y}}_{k}&= & C \widetilde{\mathbf{x}}_{k}, \\
\mathbf{e}_{k}&=& - \widetilde{\mathbf{y}}_k + \mathbf{v}_k, \\
\mathbf{u}_{k}&=& K \mathbf{e}_{k},
\end{array}
\right. 
\end{flalign}
as depicted in Fig.~\ref{kalman2}, since all the sub-systems are linear.

\section{Lower Bounds on Feedback Capacity of Parallel ACGN Channels and Recursive Coding} \label{sectionAR}

The approach we take in this paper to obtain lower bounds on the feedback capacity of parallel ACGN channels is by establishing the connection between a parallel of ACGN channels with feedback and a variant of the Kalman filter for colored Gaussian noises. Towards this end, we first present the following variant of the Kalman filter.

\subsection{A Variant of the Kalman Filter}

Consider again the Kalman filtering system given in Fig.~\ref{kalman1}. Suppose that the plant to be estimated is still given by
\begin{flalign} \label{plant3}
\left\{ \begin{array}{rcl}
\mathbf{x}_{k+1} & = & A \mathbf{x}_{k},\\
\mathbf{y}_{k} & = & C \mathbf{x}_{k} +\mathbf{v}_k,
\end{array} \right.
\end{flalign} 
only this time with an auto-regressive moving average (ARMA) colored Gaussian measurement noise $\left\{ \mathbf{v}_{k} \right\}, \mathbf{v}_{k} \in \mathbb{R}^n$ represented as
\begin{flalign} \label{ARMA}
\mathbf{v}_{k} 
&= \sum_{i=1}^{p} F_{i} \mathbf{v}_{k-i} + \widehat{\mathbf{v}}_k + \sum_{j=1}^{q} G_{j} \widehat{\mathbf{v}}_{k-j} 
, 
\end{flalign}
where $\left\{ \widehat{\mathbf{v}}_k \right\}, \widehat{\mathbf{v}}_k \in \mathbb{R}^n$ is white Gaussian with covariance $\widehat{V} = \mathbb{E} \left[ \widehat{\mathbf{v}}_{k} \widehat{\mathbf{v}}_{k}^{\mathrm{T}} \right] \succ 0$.
Equivalently, $\left\{ \mathbf{v}_k \right\}$ may be represented \cite{vaidyanathan2007theory} as the output of a linear time-invariant (LTI) filter $F \left( z \right)$ driven by input $\left\{ \widehat{\mathbf{v}}_k \right\}$, where
\begin{flalign}
F \left( z \right) 
= \left( I - \sum_{i=1}^{p} F_{i} z^{-i} \right)^{-1} \left( I + \sum_{j=1}^{q} G_{j} z^{-j} \right)
.
\end{flalign}
Herein, we assume that $F \left( z \right)$ is stable and minimum-phase.

We may now generalize the method of dealing with colored noises as employed in \cite{fang2020connection} (which in turn was developed based on \cite{anderson2012optimal}; see detailed discussions in \cite{fang2020connection}), to the case of MIMO Kalman filtering systems.


\begin{proposition} \label{anderson1}
	Suppose that $A = T \Lambda T^{-1}$, where 
	\begin{flalign}
	\Lambda = \mathrm{diag} \left( \lambda_{1}, \ldots, \lambda_{n} \right),
	\end{flalign} 
	and $\left| \lambda_{\ell} \right| \geq 1, \ell = 1, \ldots, n$.
	Denote
	\begin{flalign}
	\widehat{\mathbf{y}}_{k} 
	= - \sum_{j=1}^{q} G_{j} \widehat{\mathbf{y}}_{k-j} + \mathbf{y}_{k} - \sum_{i=1}^{p} F_{i} \mathbf{y}_{k-i}.
	\end{flalign}
	Then, \eqref{plant3} is equivalent to
	\begin{flalign} \label{plant2}
	\left\{ \begin{array}{rcl}
	\mathbf{x}_{k+1} & = & A \mathbf{x}_{k},\\
	\widehat{\mathbf{y}}_{k}  & = & \widehat{C} \mathbf{x}_{k} + \widehat{\mathbf{v}}_k,
	\end{array} \right.
	\end{flalign}
	where
	\begin{flalign} \label{c3}
	\widehat{C} 
	&= \mathrm{vec}^{-1}_{n \times n} \Bigg[ \left( T^{-\mathrm{T}} \otimes I_{n \times n} \right) \left( I_{n^2 \times n^2} - \sum_{i=1}^{p} \Lambda^{-i} \otimes F_{i} \right) \nonumber \\
	& \times \left( I_{n^2 \times n^2} + \sum_{j=1}^{q} \Lambda^{-j} \otimes G_{j} \right)^{-1} \left(T^{\mathrm{T}} \otimes I_{n \times n} \right) \mathrm{vec} \left( C \right) \Bigg].
	\end{flalign}
\end{proposition}
\begin{IEEEproof}
	Note first that since $F \left( z \right)$ is stable and minimum-phase, 
	the inverse filter
	\begin{flalign}
	F^{-1} \left( z \right)
	= \left( I - \sum_{i=1}^{p} F_{i} z^{-i} \right) \left( I + \sum_{j=1}^{q} G_{j} z^{-j} \right)^{-1} \nonumber
	\end{flalign} 
	is also stable and minimum-phase. As a result, it holds $\forall \left| z \right| \geq 1$ that
	\begin{flalign}
	\left( I - \sum_{i=1}^{p} F_{i} z^{-i} \right) \left( I + \sum_{j=1}^{q} G_{j} z^{-j} \right)^{-1} \neq 0,\nonumber
	\end{flalign}
	i.e., the region of convergence must include, though not necessarily restricted to, $\left| z \right| \geq 1$.
	Consequently, for $\left| z \right| \geq 1$, we may expand 
	\begin{flalign}
	\left( I - \sum_{i=1}^{p} F_{i} z^{-i} \right) \left( I + \sum_{j=1}^{q} G_{j} z^{-j} \right)^{-1} = I - \sum_{i=1}^{\infty} H_{i} z^{-i}
	, \nonumber 
	\end{flalign}
	and thus $\left\{ \widehat{\mathbf{v}}_k \right\}$ can be reconstructed from $\left\{ \mathbf{v}_k \right\}$ as \cite{vaidyanathan2007theory} 
	\begin{flalign} 
	\widehat{\mathbf{v}}_{k}
	=  \mathbf{v}_{k} - \sum_{i=1}^{\infty} H_{i} \mathbf{v}_{k-i}
	= - \sum_{j=1}^{q} G_{j} \widehat{\mathbf{v}}_{k-j} + \mathbf{v}_{k} - \sum_{i=1}^{p} F_{i} \mathbf{v}_{k-i} 
	.  \nonumber
	\end{flalign}
	Accordingly, we may also rewrite
	\begin{flalign}
	\widehat{\mathbf{y}}_{k} 
	&= 	- \sum_{j=1}^{q} G_{j} \widehat{\mathbf{y}}_{k-j} + \mathbf{y}_{k} - \sum_{i=1}^{p} F_{i} \mathbf{y}_{k-i}
	=  \mathbf{y}_{k} - \sum_{i=1}^{\infty} H_{i} \mathbf{y}_{k-i} \nonumber \\
	&=  \mathbf{y}_{k} - \sum_{i=1}^{\infty} H_{i} \left( C \mathbf{x}_{k-i} +\mathbf{v}_{k-i} \right) \nonumber \\
	&=  C \mathbf{x}_{k} - \sum_{i=1}^{\infty} H_{i} C \mathbf{x}_{k-i} + \mathbf{v}_{k} - \sum_{i=1}^{\infty} H_{i} \mathbf{v}_{k-i} \nonumber \\
	&=  C \mathbf{x}_{k}- \sum_{i=1}^{\infty} H_{i} C \mathbf{x}_{k-i} + \widehat{\mathbf{v}}_k. \nonumber
	\end{flalign}
	Meanwhile, since $A$ is anti-stable (and thus invertible), we have $\mathbf{x}_{k-i} = A^{-i} \mathbf{x}_{k}$. As a result, 
	\begin{flalign}
	\widehat{\mathbf{y}}_{k} 
	&=  C \mathbf{x}_{k} - \sum_{i=1}^{\infty} H_{i} C \mathbf{x}_{k-i} + \widehat{\mathbf{v}}_k \nonumber \\
	&=  \left( C - \sum_{i=1}^{\infty} H_{i} C A^{-i} \right) \mathbf{x}_{k} + \widehat{\mathbf{v}}_k. \nonumber
	\end{flalign}
	In addition,
	\begin{flalign}
	&\mathrm{vec} \left( C - \sum_{i=1}^{\infty} H_{i} C A^{-i} \right)
	= \mathrm{vec} \left( C \right) - \mathrm{vec} \left( \sum_{i=1}^{\infty} H_{i} C A^{-i} \right)
	\nonumber \\
	&~~~~ = \mathrm{vec} \left( C \right) - \sum_{i=1}^{\infty} \left[ \left( A^{-i} \right)^{\mathrm{T}} \otimes H_{i} \right] \mathrm{vec} \left( C \right) \nonumber \\
	&~~~~ = \left[ I_{n^2 \times n^2} - \sum_{i=1}^{\infty} \left( A^{-i} \right)^{\mathrm{T}} \otimes H_{i} \right] \mathrm{vec} \left( C \right), \nonumber
	\end{flalign} 
	and hence
	\begin{flalign}
	&C - \sum_{i=1}^{\infty} H_{i} C A^{-i}
	\nonumber \\
	&~~~~ = \mathrm{vec}^{-1}_{n \times n} \left\{ \left[ I_{n^2 \times n^2} - \sum_{i=1}^{\infty} \left( A^{-i} \right)^{\mathrm{T}} \otimes H_{i} \right] \mathrm{vec} \left( C \right) \right\}. \nonumber
	\end{flalign} 
	
	Note then that
	\begin{flalign}
	&I_{n^2 \times n^2} - \sum_{i=1}^{\infty} \left( A^{-i} \right)^{\mathrm{T}} \otimes H_{i} \nonumber \\
	& = I_{n^2 \times n^2} - \sum_{i=1}^{\infty} \left[ \left( T \Lambda T^{-1} \right)^{-i} \right]^{\mathrm{T}} \otimes H_{i} \nonumber \\
	& = I_{n^2 \times n^2} - \sum_{i=1}^{\infty} \left( T \Lambda^{-i} T^{-1} \right)^{\mathrm{T}} \otimes \left( I_{n \times n} H_{i} I_{n \times n} \right) \nonumber \\
	& = I_{n^2 \times n^2} - \sum_{i=1}^{\infty} \left( T^{-\mathrm{T}} \Lambda^{-i} T^{\mathrm{T}} \right) \otimes \left( I_{n \times n} H_{i} I_{n \times n} \right) \nonumber \\
	& = I_{n^2 \times n^2} - \sum_{i=1}^{\infty} \left( T^{-\mathrm{T}} \otimes I_{n \times n} \right) \left( \Lambda^{-i} \otimes H_{i} \right) \left(T^{\mathrm{T}} \otimes I_{n \times n} \right) \nonumber \\
	& = I_{n^2 \times n^2} - \left( T^{-\mathrm{T}} \otimes I_{n \times n} \right) \left( \sum_{i=1}^{\infty} \Lambda^{-i} \otimes H_{i} \right) \left(T^{\mathrm{T}} \otimes I_{n \times n} \right)
	. \nonumber
	\end{flalign} 
	Moreover, since $T^{\mathrm{T}}$ is invertible and the eigenvalues of $T^{\mathrm{T}} \otimes I_{n \times n}$ are given by the $n$ copies of the eigenvalues of $T^{\mathrm{T}}$, it thus follows that $T^{\mathrm{T}} \otimes I_{n \times n}$ is invertible and 
	\begin{flalign}
	I_{n^2 \times n^2} 
	&= \left( T^{\mathrm{T}} \otimes I_{n \times n} \right)^{-1} \left(T^{\mathrm{T}} \otimes I_{n \times n} \right) \nonumber \\
	&= \left( T^{-\mathrm{T}} \otimes I_{n \times n} \right) \left(T^{\mathrm{T}} \otimes I_{n \times n} \right)
	. \nonumber
	\end{flalign} 
	Accordingly, 
	\begin{flalign}
	&I_{n^2 \times n^2} - \left( T^{-\mathrm{T}} \otimes I_{n \times n} \right) \left( \sum_{i=1}^{\infty} \Lambda^{-i} \otimes H_{i} \right) \left(T^{\mathrm{T}} \otimes I_{n \times n} \right) \nonumber \\
	&= \left( T^{-\mathrm{T}} \otimes I_{n \times n} \right) \left(T^{\mathrm{T}} \otimes I_{n \times n} \right) \nonumber \\
	&~~~~ - \left( T^{-\mathrm{T}} \otimes I_{n \times n} \right) \left( \sum_{i=1}^{\infty} \Lambda^{-i} \otimes H_{i} \right) \left(T^{\mathrm{T}} \otimes I_{n \times n} \right) \nonumber \\
	&= \left( T^{-\mathrm{T}} \otimes I_{n \times n} \right) \left( I_{n^2 \times n^2} - \sum_{i=1}^{\infty} \Lambda^{-i} \otimes H_{i} \right) \left(T^{\mathrm{T}} \otimes I_{n \times n} \right)
	. \nonumber
	\end{flalign} 
	In addition,
	\begin{flalign}
	& I_{n^2 \times n^2} - \sum_{i=1}^{\infty} \Lambda^{-i} \otimes H_{i} \nonumber \\
	& = I_{n^2 \times n^2} - \sum_{i=1}^{\infty} \left[
	\begin{array}{cccc}
	\lambda_{1}^{-i} & \cdots & 0\\
	\vdots & \ddots & \vdots\\
	0  & \cdots & \lambda_{n}^{-i}\\
	\end{array} \right] \otimes H_{i} \nonumber \\
	& = I_{n^2 \times n^2} - \sum_{i=1}^{\infty} \left[
	\begin{array}{cccc}
	\lambda_{1}^{-i} H_{i} & \cdots & 0\\
	\vdots & \ddots & \vdots\\
	0 & \cdots & \lambda_{n}^{-i} H_{i} \\
	\end{array} \right] \nonumber \\
	& = \left[
	\begin{array}{ccc}
	I_{n \times n} - \sum_{i=1}^{\infty} \lambda_{1}^{-i} H_{i} & \cdots & 0\\
	\vdots & \ddots & \vdots\\
	0 & \cdots & I_{n \times n} - \sum_{i=1}^{\infty} \lambda_{n}^{-i} H_{i} \\
	\end{array} \right]
	. \nonumber
	\end{flalign} 
	Meanwhile, we have already shown that $\forall \left| z \right| \geq 1$,
	\begin{flalign}
	&I_{n \times n} - \sum_{i=1}^{\infty} H_{i} z^{-i} \nonumber \\
	&~~~~ = \left( I_{n \times n} - \sum_{i=1}^{p} F_{i} z^{-i} \right) \left( I_{n \times n} + \sum_{j=1}^{q} G_{j} z^{-j} \right)^{-1},
	\nonumber
	\end{flalign}
	i.e., $I_{n \times n} - \sum_{i=1}^{\infty} H_{i} z^{-i}$ converges.
	As such, since $\left| \lambda_{\ell} \right| \geq 1, \ell = 1, \ldots, n$, we have
	\begin{flalign}
	&I_{n \times n} - \sum_{i=1}^{\infty} \lambda_{\ell}^{-i} H_{i} 
	= I_{n \times n} - \sum_{i=1}^{\infty} H_{i} \lambda_{\ell}^{-i} \nonumber \\
	&= \left( I_{n \times n} - \sum_{i=1}^{p} F_{i} \lambda_{\ell}^{-i} \right) \left( I_{n \times n} + \sum_{j=1}^{q} G_{j} \lambda_{\ell}^{-j} \right)^{-1} \nonumber \\
	&= \left( I_{n \times n} - \sum_{i=1}^{p} \lambda_{\ell}^{-i} F_{i} \right) \left( I_{n \times n} + \sum_{j=1}^{q} \lambda_{\ell}^{-j} G_{j} \right)^{-1}.
	\nonumber
	\end{flalign}
	Therefore,
	\begin{flalign}
	& \left[
	\begin{array}{ccc}
	I_{n \times n} - \sum_{i=1}^{\infty} \lambda_{1}^{-i} H_{i} & \cdots & 0\\
	\vdots & \ddots & \vdots\\
	0 & \cdots & I_{n \times n} - \sum_{i=1}^{\infty} \lambda_{n}^{-i} H_{i} \\
	\end{array} \right] \nonumber \\
	& = \left[
	\begin{array}{ccc}
	I_{n \times n} - \sum_{i=1}^{p} \lambda_{1}^{-i} F_{i} & \cdots & 0\\
	\vdots & \ddots & \vdots\\
	0 & \cdots & I_{n \times n} - \sum_{i=1}^{p} \lambda_{n}^{-i} F_{i} \\
	\end{array} \right] \nonumber \\
	&\times \left[
	\begin{array}{ccc}
	I_{n \times n} + \sum_{j=1}^{q} \lambda_{1}^{-j} G_{j} & \cdots & 0\\
	\vdots & \ddots & \vdots\\
	0 & \cdots & I_{n \times n} + \sum_{j=1}^{q} \lambda_{n}^{-j} G_{j} \\
	\end{array} \right]^{-1} \nonumber \\
	& = \left( I_{n^2 \times n^2} - \sum_{i=1}^{p} \left[
	\begin{array}{cccc}
	\lambda_{1}^{-i} F_{i}  & \cdots & 0\\
	\vdots & \ddots & \vdots\\
	0 & \cdots & \lambda_{n}^{-i} F_{i} \\
	\end{array} \right] \right) \nonumber \\
	&~~~~ \times \left( I_{n^2 \times n^2} + \sum_{j=1}^{q} \left[
	\begin{array}{cccc}
	\lambda_{1}^{-j} G_{j} & \cdots & 0\\
	\vdots & \ddots & \vdots\\
	0 & \cdots & \lambda_{n}^{-j} G_{j} \\
	\end{array} \right] \right)^{-1} \nonumber \\
	& = \left( I_{n^2 \times n^2} - \sum_{i=1}^{p} \left[
	\begin{array}{cccc}
	\lambda_{1}^{-i} & \cdots & 0\\
	\vdots & \ddots & \vdots\\
	0  & \cdots & \lambda_{n}^{-i}\\
	\end{array} \right] \otimes F_{i} \right) \nonumber \\
	&~~~~ \times \left( I_{n^2 \times n^2} + \sum_{j=1}^{q} \left[
	\begin{array}{cccc}
	\lambda_{1}^{-j} & \cdots & 0\\
	\vdots & \ddots & \vdots\\
	0  & \cdots & \lambda_{n}^{-j}\\
	\end{array} \right] \otimes G_{j} \right)^{-1} \nonumber \\
	& = \left( I_{n^2 \times n^2} - \sum_{i=1}^{p} \Lambda^{-i} \otimes F_{i} \right) \left( I_{n^2 \times n^2} + \sum_{j=1}^{q} \Lambda^{-j} \otimes G_{j} \right)^{-1}
	. \nonumber
	\end{flalign} 	
	As a result,
	\begin{flalign}
	&I_{n^2 \times n^2} - \sum_{i=1}^{\infty} \left( A^{-i} \right)^{\mathrm{T}} \otimes H_{i} \nonumber \\
	&= \left( T^{-\mathrm{T}} \otimes I_{n \times n} \right) \left( I_{n^2 \times n^2} - \sum_{i=1}^{\infty} \Lambda^{-i} \otimes H_{i} \right) \left(T^{\mathrm{T}} \otimes I_{n \times n} \right) \nonumber \\
	&= \left( T^{-\mathrm{T}} \otimes I_{n \times n} \right) \left( I_{n^2 \times n^2} - \sum_{i=1}^{p} \Lambda^{-i} \otimes F_{i} \right) \nonumber \\
	&~~~~ \times \left( I_{n^2 \times n^2} + \sum_{j=1}^{q} \Lambda^{-j} \otimes G_{j} \right)^{-1} \left(T^{\mathrm{T}} \otimes I_{n \times n} \right)
	. \nonumber
	\end{flalign} 
	
	To sum it up, we have
	\begin{flalign}
	& C - \sum_{i=1}^{\infty} H_{i} C A^{-i} \nonumber \\
	& = \mathrm{vec}^{-1}_{n \times n} \left\{ \left[ I_{n^2 \times n^2} - \sum_{i=1}^{\infty} \left( A^{-i} \right)^{\mathrm{T}} \otimes H_{i} \right] \mathrm{vec} \left( C \right) \right\} \nonumber \\
	& = \mathrm{vec}^{-1}_{n \times n} \Bigg[ \left( T^{-\mathrm{T}} \otimes I_{n \times n} \right) \left( I_{n^2 \times n^2} - \sum_{i=1}^{p} \Lambda^{-i} \otimes F_{i} \right) \nonumber \\
	&~~~~ \times \left( I_{n^2 \times n^2} + \sum_{j=1}^{q} \Lambda^{-j} \otimes G_{j} \right)^{-1} \left(T^{\mathrm{T}} \otimes I_{n \times n} \right) \mathrm{vec} \left( C \right) \Bigg]
	. \nonumber
	\end{flalign} 
	This completes the proof.
\end{IEEEproof}

Meanwhile, the Kalman filter for \eqref{plant2} is given by
\begin{flalign} \label{estimator2}
\left\{ \begin{array}{rcl}
\overline{\mathbf{x}}_{k+1}&=&A \overline{\mathbf{x}}_{k} +\mathbf{u}_k, \\
\overline{\mathbf{y}}_{k}&= & \widehat{C} \overline{\mathbf{x}}_{k}, \\
\mathbf{e}_{k}&=& \widehat{\mathbf{y}}_{k}-\overline{\mathbf{y}}_k, \\
\mathbf{u}_{k}&=& \widehat{K}_{k} \mathbf{e}_{k},
\end{array}
\right.
\end{flalign}
where $\overline{\mathbf{x}}_{k} \in \mathbb{R}^n$, $\overline{\mathbf{y}}_{k} \in \mathbb{R}^n$, $\mathbf{e}_{k} \in \mathbb{R}^n$, and $\mathbf{\mathbf{u}}_{k} \in \mathbb{R}^n$. 
\begin{figure}
	\begin{center}
		\vspace{-3mm}
		\includegraphics [width=0.5\textwidth]{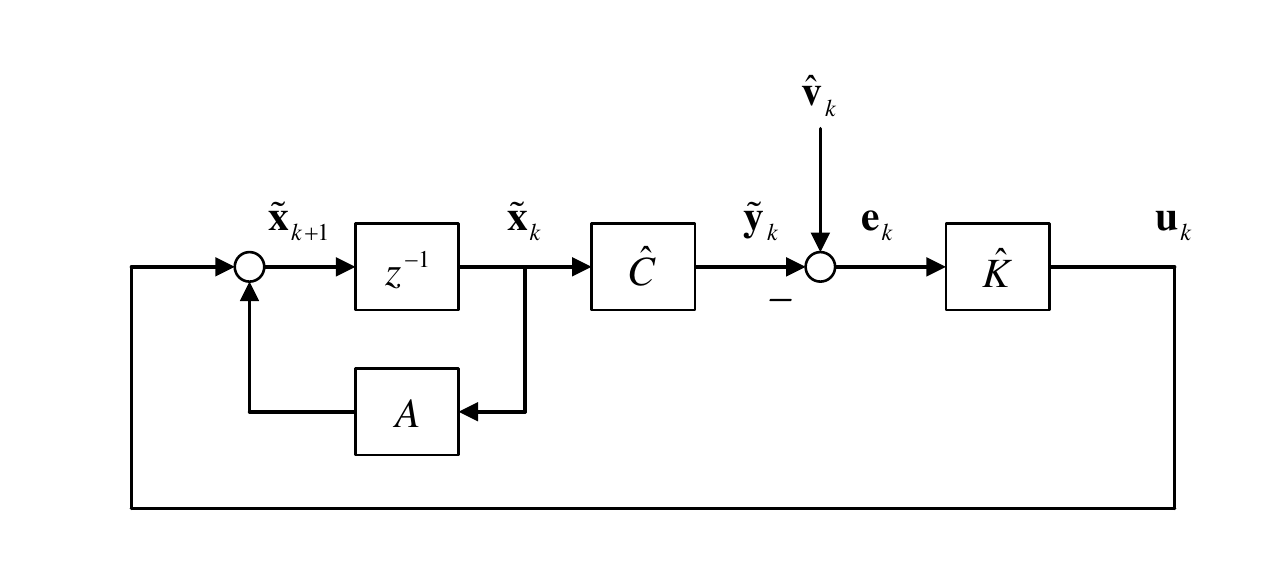}
		\vspace{-6mm}
		\caption{The steady-state integrated Kalman filter for colored noises.}
		\label{kalman3}
	\end{center}
	\vspace{-3mm}
\end{figure}
Furthermore, when $\left( A, \widehat{C} \right)$ is detectable, the Kalman filtering system converges, i.e., the state estimation error $\left\{ \mathbf{x}_{k} - \overline{\mathbf{x}}_{k} \right\}$ is asymptotically stationary. Moreover, in steady state, the optimal state estimation error covariance
\begin{flalign}
P =\lim_{k\to \infty} \mathbb{E} \left[ \left( \mathbf{x}_{k} - \overline{\mathbf{x}}_{k} \right) \left( \mathbf{x}_{k} - \overline{\mathbf{x}}_{k} \right)^{\mathrm{T}} \right] \nonumber
\end{flalign}
attained by the Kalman filter is given by the (non-zero) positive semi-definite solution to the algebraic Riccati equation
\begin{flalign} \label{are2}
P
= A P A^{\mathrm{T}} - A P \widehat{C}^{\mathrm{T}} \left( \widehat{C} P \widehat{C}^{\mathrm{T}} + \widehat{V} \right)^{-1} \widehat{C} P A^{\mathrm{T}}, 
\end{flalign} 
whereas the steady-state observer gain is given by 
\begin{flalign} \label{gain2}
\widehat{K} = A P \widehat{C}^{\mathrm{T}} \left( \widehat{C} P \widehat{C}^{\mathrm{T}} + \widehat{V} \right)^{-1}.
\end{flalign}  
Again, by letting $\widetilde{\mathbf{x}}_{k} = \overline{\mathbf{x}}_{k} - \mathbf{x}_{k} $ and $\widetilde{\mathbf{y}}_{k} = \overline{\mathbf{y}}_{k} - \widehat{\mathbf{z}}_{k} = \overline{\mathbf{y}}_{k} - \widehat{C} \mathbf{x}_{k} $, we may integrate the steady-state systems of \eqref{plant2} and \eqref{estimator2} into
\begin{flalign} \label{integrate}
\left\{ \begin{array}{rcl}
\widetilde{\mathbf{x}}_{k+1}&=&A \widetilde{\mathbf{x}}_{k} +\mathbf{u}_k, \\
\widetilde{\mathbf{y}}_{k}&= & \widehat{C} \widetilde{\mathbf{x}}_{k}, \\
\mathbf{e}_{k}&=& - \widetilde{\mathbf{y}}_k + \widehat{\mathbf{v}}_k, \\
\mathbf{u}_{k}&=& \widehat{K} \mathbf{e}_{k},
\end{array}
\right.
\end{flalign}
as depicted in Fig.~\ref{kalman3}. In addition, it may be verified that the closed-loop system given in \eqref{integrate} and Fig.~\ref{kalman3} is stable \cite{anderson2012optimal, astrom2010feedback}.

%

In fact, we may design matrix $C$ specifically to render matrix $\widehat{C}$ an identity matrix.

\begin{proposition} \label{designC}
	Suppose that $A = T \Lambda T^{-1}$, where 
	\begin{flalign}
	\Lambda = \mathrm{diag} \left( \lambda_{1}, \ldots, \lambda_{n} \right),
	\end{flalign} 
	and $\left| \lambda_{\ell} \right| \geq 1, \ell = 1, \ldots, n$.
	If 
	\begin{flalign} \label{C}
	&C = \mathrm{vec}^{-1}_{n \times n} \Bigg[ \left( T^{-\mathrm{T}} \otimes I_{n \times n} \right) \left( I_{n^2 \times n^2} + \sum_{j=1}^{q} \Lambda^{-j} \otimes G_{j} \right) \nonumber \\
	& \times \left( I_{n^2 \times n^2} - \sum_{i=1}^{p} \Lambda^{-i} \otimes F_{i} \right)^{-1} \left(T^{\mathrm{T}} \otimes I_{n \times n} \right) \mathrm{vec} \left( I_{n \times n} \right) \Bigg]
	,
	\end{flalign} 
	then
	\begin{flalign} 
	\widehat{C} 
	= I_{n \times n}
	.
	\end{flalign}
\end{proposition}

\begin{IEEEproof}
	It is known from the proof of Proposition~\ref{anderson1} that
	\begin{flalign}
	&\left( T^{-\mathrm{T}} \otimes I_{n \times n} \right) \left( I_{n^2 \times n^2} - \sum_{i=1}^{p} \Lambda^{-i} \otimes F_{i} \right) \nonumber \\
	&~~~~ \times \left( I_{n^2 \times n^2} + \sum_{j=1}^{q} \Lambda^{-j} \otimes G_{j} \right)^{-1} \left(T^{\mathrm{T}} \otimes I_{n \times n} \right)  \nonumber
	\end{flalign} 
	is invertible, and it can be verified that its inverse is given by
	\begin{flalign}
	&\left( T^{-\mathrm{T}} \otimes I_{n \times n} \right) \left( I_{n^2 \times n^2} + \sum_{j=1}^{q} \Lambda^{-j} \otimes G_{j} \right) \nonumber \\
	&~~~~ \times \left( I_{n^2 \times n^2} - \sum_{i=1}^{p} \Lambda^{-i} \otimes F_{i} \right)^{-1} \left(T^{\mathrm{T}} \otimes I_{n \times n} \right)
	. \nonumber
	\end{flalign} 
	Hence, if \eqref{C} holds, then
	\begin{flalign} 
	\widehat{C} 
	&= \mathrm{vec}^{-1}_{n \times n} \Bigg[ \left( T^{-\mathrm{T}} \otimes I_{n \times n} \right) \left( I_{n^2 \times n^2} - \sum_{i=1}^{p} \Lambda^{-i} \otimes F_{i} \right) \nonumber \\
	& \times \left( I_{n^2 \times n^2} + \sum_{j=1}^{q} \Lambda^{-j} \otimes G_{j} \right)^{-1} \left(T^{\mathrm{T}} \otimes I_{n \times n} \right) \mathrm{vec} \left( C \right) \Bigg] \nonumber \\
	&= \mathrm{vec}^{-1}_{n \times n} \Bigg[ \left( T^{-\mathrm{T}} \otimes I_{n \times n} \right) \left( I_{n^2 \times n^2} - \sum_{i=1}^{p} \Lambda^{-i} \otimes F_{i} \right) \nonumber \\
	& \times \left( I_{n^2 \times n^2} + \sum_{j=1}^{q} \Lambda^{-j} \otimes G_{j} \right)^{-1} \left(T^{\mathrm{T}} \otimes I_{n \times n} \right) \nonumber \\
	& \times \left( T^{-\mathrm{T}} \otimes I_{n \times n} \right) \left( I_{n^2 \times n^2} + \sum_{j=1}^{q} \Lambda^{-j} \otimes G_{j} \right) \nonumber \\
	& \times \left( I_{n^2 \times n^2} - \sum_{i=1}^{p} \Lambda^{-i} \otimes F_{i} \right)^{-1} \left(T^{\mathrm{T}} \otimes I_{n \times n} \right) \mathrm{vec} \left( I_{n \times n} \right) \Bigg] \nonumber \\
	& = I_{n \times n}
	. \nonumber
	\end{flalign}
	This completes the proof.
\end{IEEEproof}

Note that when $\widehat{C} = I_{n \times n}$, the pair $\left( A, \widehat{C} \right)$ is always observable (and thus always detectable \cite{astrom2010feedback}). To see this, note that if $\widehat{C} = I_{n \times n}$, then the observation matrix for $\left( A, \widehat{C} \right)$ becomes
\begin{flalign}
&\left[
\begin{array}{c}
\widehat{C} \\
\widehat{C} A \\
\widehat{C} A^2 \\
\vdots \\
\widehat{C} A^{n-1}\\
\end{array} \right] 
= \left[
\begin{array}{c}
I_{n \times n} \\
A \\
A^2 \\
\vdots \\
A^{n-1}\\
\end{array} \right],
\end{flalign} 
which has a row rank of $n$, indicating that $\left( A, \widehat{C} \right)$ is observable \cite{astrom2010feedback}, regardless of what $A$ is.
As such, in this case the Kalman filtering system always converges, whereas \eqref{are2} reduces to
\begin{flalign} \label{are3}
P
= A P A^{\mathrm{T}} - A P \left( P + \widehat{V} \right)^{-1} P A^{\mathrm{T}},
\end{flalign} 
and \eqref{gain2} reduces to
\begin{flalign}
\widehat{K} = A P \left( P + \widehat{V} \right)^{-1}.
\end{flalign}

\subsection{Feedback Capacity of Parallel ACGN Channels}

We now proceed to obtain lower bounds on feedback capacity as well as the corresponding recursive coding schemes, based upon the discussions in the previous sub-section. We first examine the solution to the algebraic Riccati equation given by \eqref{are2} when $A$ and $C$ are designed specifically.

\begin{theorem} \label{t1}
	Suppose that in \eqref{ARMA}, $\widehat{V} = U_{\widehat{\mathbf{v}}} \Lambda_{\widehat{\mathbf{v}}} U^{\mathrm{T}}_{\widehat{\mathbf{v}}}$, where 
	\begin{flalign} 
	\Lambda_{\widehat{\mathbf{v}}} = \mathrm{diag} \left( \widehat{V}_1, \ldots, \widehat{V}_{n}\right),
	\end{flalign}
	and $U_{\widehat{\mathbf{v}}}$ is an orthogonal matrix.
	If
	\begin{flalign}
	A 
	= 
	U_{\widehat{\mathbf{v}}} \Lambda U^{\mathrm{T}}_{\widehat{\mathbf{v}}},
	\end{flalign}
	and 
	\begin{flalign} \label{Cdesign2}
	&C = \mathrm{vec}^{-1}_{n \times n} \Bigg[ \left( U_{\widehat{\mathbf{v}}}^{-\mathrm{T}} \otimes I_{n \times n} \right) \left( I_{n^2 \times n^2} + \sum_{j=1}^{q} \Lambda^{-j} \otimes G_{j} \right) \nonumber \\
	& \times \left( I_{n^2 \times n^2} - \sum_{i=1}^{p} \Lambda^{-i} \otimes F_{i} \right)^{-1} \left(U_{\widehat{\mathbf{v}}}^{\mathrm{T}} \otimes I_{n \times n} \right) \mathrm{vec} \left( I_{n \times n} \right) \Bigg]
	,
	\end{flalign} 
	where
	\begin{flalign}
	\Lambda
	= \pm \left( \Lambda_{\widetilde{\mathbf{x}}} + \Lambda_{\widehat{\mathbf{v}}} \right)^{\frac{1}{2}} \Lambda_{\widehat{\mathbf{v}}}^{-\frac{1}{2}},
	\end{flalign}
	and
	\begin{flalign} 
	\Lambda_{\widetilde{\mathbf{x}}} = \mathrm{diag} \left( P_1, \ldots, P_{n} \right) \succeq 0,~\Lambda_{\widetilde{\mathbf{x}}} \neq 0,
	\end{flalign}
	then the (non-zero) positive semi-definite solution to \eqref{are2} is given by
	\begin{flalign} 
	P = U_{\widehat{\mathbf{v}}} \Lambda_{\widetilde{\mathbf{x}}} U^{\mathrm{T}}_{\widehat{\mathbf{v}}}.
	\end{flalign}
\end{theorem}

\begin{IEEEproof}
	Note first that the eigenvalues of 
	\begin{flalign}
	A = U_{\widehat{\mathbf{v}}} \Lambda U^{\mathrm{T}}_{\widehat{\mathbf{v}}} 
	= \pm U_{\widehat{\mathbf{v}}} \left( \Lambda_{\widetilde{\mathbf{x}}} + \Lambda_{\widehat{\mathbf{v}}} \right)^{\frac{1}{2}} \Lambda_{\widehat{\mathbf{v}}}^{-\frac{1}{2}} U^{\mathrm{T}}_{\widehat{\mathbf{v}}} \nonumber
	\end{flalign}
	are given by
	\begin{flalign}
	\lambda_{\ell} =  \sqrt{\frac{ P_{\ell} + \widehat{V}_{\ell} }{ \widehat{V}_{\ell} }},~\ell = 1, \ldots, n, \nonumber
	\end{flalign} 
	or 
	\begin{flalign}
	\lambda_{\ell} = - \sqrt{\frac{ P_{\ell} + \widehat{V}_{\ell} }{ \widehat{V}_{\ell} }},~\ell = 1, \ldots, n. \nonumber
	\end{flalign} 
	Then, since
	\begin{flalign} 
	\mathrm{diag} \left( \widehat{V}_1, \ldots, \widehat{V}_{n}\right) \succ 0, \nonumber
	\end{flalign}
	and
	\begin{flalign} 
	\mathrm{diag} \left( P_1, \ldots, P_{n} \right) \succeq 0, \nonumber
	\end{flalign}
	we have
	\begin{flalign}
	\left| \lambda_{\ell} \right| = \left| \sqrt{\frac{ P_{\ell} + \widehat{V}_{\ell} }{ \widehat{V}_{\ell} }} \right| \geq 1,~\ell = 1, \ldots, n. \nonumber
	\end{flalign} 
	Therefore, according to Proposition~\ref{anderson1} and Proposition~\ref{designC}, when $C$ is given by \eqref{Cdesign2},
	it follows that $\widehat{C} = I_{n \times n}$ while \eqref{are3} holds.
	On the other hand, it may be verified that $P = U_{\widehat{\mathbf{v}}} \Lambda_{\widetilde{\mathbf{x}}} U^{\mathrm{T}}_{\widehat{\mathbf{v}}}$ is the (non-zero) positive semi-definite solution to \eqref{are3} as
	\begin{flalign}
	&U_{\widehat{\mathbf{v}}} \Lambda_{\widetilde{\mathbf{x}}} U^{\mathrm{T}}_{\widehat{\mathbf{v}}}
	- A U_{\widehat{\mathbf{v}}} \Lambda_{\widetilde{\mathbf{x}}} U^{\mathrm{T}}_{\widehat{\mathbf{v}}} A^{\mathrm{T}} \nonumber \\
	&~~~~ + A U_{\widehat{\mathbf{v}}} \Lambda_{\widetilde{\mathbf{x}}} U^{\mathrm{T}}_{\widehat{\mathbf{v}}} \left( U_{\widehat{\mathbf{v}}} \Lambda_{\widetilde{\mathbf{x}}} U^{\mathrm{T}}_{\widehat{\mathbf{v}}} +  \widehat{V} \right)^{-1} U_{\widehat{\mathbf{v}}} \Lambda_{\widetilde{\mathbf{x}}} U^{\mathrm{T}}_{\widehat{\mathbf{v}}} A^{\mathrm{T}} \nonumber \\
	&= U_{\widehat{\mathbf{v}}} \Lambda_{\widetilde{\mathbf{x}}} U^{\mathrm{T}}_{\widehat{\mathbf{v}}}
	- U_{\widehat{\mathbf{v}}} \Lambda U^{\mathrm{T}}_{\widehat{\mathbf{v}}} U_{\widehat{\mathbf{v}}} \Lambda_{\widetilde{\mathbf{x}}} U^{\mathrm{T}}_{\widehat{\mathbf{v}}} U_{\widehat{\mathbf{v}}} \Lambda U^{\mathrm{T}}_{\widehat{\mathbf{v}}} \nonumber \\
	&~~~~ + U_{\widehat{\mathbf{v}}} \Lambda U^{\mathrm{T}}_{\widehat{\mathbf{v}}} U_{\widehat{\mathbf{v}}} \Lambda_{\widetilde{\mathbf{x}}} U^{\mathrm{T}}_{\widehat{\mathbf{v}}} \left( U_{\widehat{\mathbf{v}}} \Lambda_{\widetilde{\mathbf{x}}} U^{\mathrm{T}}_{\widehat{\mathbf{v}}} +  U_{\widehat{\mathbf{v}}} \Lambda_{\widehat{\mathbf{v}}} U^{\mathrm{T}}_{\widehat{\mathbf{v}}} \right)^{-1} \nonumber \\
	&~~~~~~~~ \times U_{\widehat{\mathbf{v}}} \Lambda_{\widetilde{\mathbf{x}}} U^{\mathrm{T}}_{\widehat{\mathbf{v}}} U_{\widehat{\mathbf{v}}} \Lambda U^{\mathrm{T}}_{\widehat{\mathbf{v}}} \nonumber \\
	&= U_{\widehat{\mathbf{v}}} \Lambda_{\widetilde{\mathbf{x}}} U^{\mathrm{T}}_{\widehat{\mathbf{v}}}
	- U_{\widehat{\mathbf{v}}} \Lambda \Lambda_{\widetilde{\mathbf{x}}} \Lambda U^{\mathrm{T}}_{\widehat{\mathbf{v}}} + U_{\widehat{\mathbf{v}}} \Lambda  \Lambda_{\widetilde{\mathbf{x}}} \left( \Lambda_{\widetilde{\mathbf{x}}} + \Lambda_{\widehat{\mathbf{v}}} \right)^{-1}  \Lambda_{\widetilde{\mathbf{x}}} \Lambda U^{\mathrm{T}}_{\widehat{\mathbf{v}}} \nonumber \\
	&= U_{\widehat{\mathbf{v}}} \left[ \Lambda_{\widetilde{\mathbf{x}}}
	- \Lambda \Lambda_{\widetilde{\mathbf{x}}} \Lambda + \Lambda  \Lambda_{\widetilde{\mathbf{x}}} \left( \Lambda_{\widetilde{\mathbf{x}}} + \Lambda_{\widehat{\mathbf{v}}} \right)^{-1}  \Lambda_{\widetilde{\mathbf{x}}} \Lambda \right] U^{\mathrm{T}}_{\widehat{\mathbf{v}}} \nonumber \\
	&= U_{\widehat{\mathbf{v}}} \Big[ \Lambda_{\widetilde{\mathbf{x}}}
	- \Lambda \Lambda_{\widetilde{\mathbf{x}}} \left( \Lambda_{\widetilde{\mathbf{x}}} + \Lambda_{\widehat{\mathbf{v}}} \right)^{-1} \left( \Lambda_{\widetilde{\mathbf{x}}} + \Lambda_{\widehat{\mathbf{v}}} \right) \Lambda \nonumber \\
	&~~~~ + \Lambda  \Lambda_{\widetilde{\mathbf{x}}} \left( \Lambda_{\widetilde{\mathbf{x}}} + \Lambda_{\widehat{\mathbf{v}}} \right)^{-1}  \Lambda_{\widetilde{\mathbf{x}}} \Lambda \Big] U^{\mathrm{T}}_{\widehat{\mathbf{v}}} \nonumber \\
	&= U_{\widehat{\mathbf{v}}} \Big[ \Lambda_{\widetilde{\mathbf{x}}}
	- \Lambda \Lambda_{\widetilde{\mathbf{x}}} \left( \Lambda_{\widetilde{\mathbf{x}}} + \Lambda_{\widehat{\mathbf{v}}} \right)^{-1} \Lambda_{\widehat{\mathbf{v}}} \Lambda  \Big] U^{\mathrm{T}}_{\widehat{\mathbf{v}}} \nonumber \\
	&= U_{\widehat{\mathbf{v}}} \Big[ \Lambda_{\widetilde{\mathbf{x}}}
	- \Lambda_{\widetilde{\mathbf{x}}} \left( \Lambda_{\widetilde{\mathbf{x}}} + \Lambda_{\widehat{\mathbf{v}}} \right)^{-1} \Lambda_{\widehat{\mathbf{v}}} \Lambda^2  \Big] U^{\mathrm{T}}_{\widehat{\mathbf{v}}} \nonumber \\
	&= U_{\widehat{\mathbf{v}}} \Big[ \Lambda_{\widetilde{\mathbf{x}}}
	- \Lambda_{\widetilde{\mathbf{x}}} \left( \Lambda_{\widetilde{\mathbf{x}}} + \Lambda_{\widehat{\mathbf{v}}} \right)^{-1} \Lambda_{\widehat{\mathbf{v}}}  \left( \Lambda_{\widetilde{\mathbf{x}}} + \Lambda_{\widehat{\mathbf{v}}} \right) \Lambda_{\widehat{\mathbf{v}}}^{-1}  \Big] U^{\mathrm{T}}_{\widehat{\mathbf{v}}} \nonumber \\
	&= U_{\widehat{\mathbf{v}}} \Big[ \Lambda_{\widetilde{\mathbf{x}}}
	- \Lambda_{\widetilde{\mathbf{x}}} \Big] U^{\mathrm{T}}_{\widehat{\mathbf{v}}} = 0
	. \nonumber
	\end{flalign} 
	(Note also that clearly $P = 0$ is the other positive semi-definite solution to \eqref{are3}, which is not relevant herein though.) This completes the proof.
\end{IEEEproof}


Note that herein $\Lambda$ and $P$ can respectively be rewritten as
\begin{flalign}
\Lambda
= \pm  \left[
\begin{array}{cccc}
\sqrt{1 + \frac{P_1}{\widehat{V}_1}} & \cdots & 0\\
\vdots & \ddots & \vdots\\
0 & \cdots & \sqrt{1 + \frac{P_n}{\widehat{V}_n}}\\
\end{array} \right],
\end{flalign}
and
\begin{flalign}
P = U_{\widehat{\mathbf{v}}} \left[
\begin{array}{cccc}
P_1 & \cdots & 0\\
\vdots & \ddots & \vdots\\
0 & \cdots & P_n\\
\end{array} \right] U^{\mathrm{T}}_{\widehat{\mathbf{v}}}.
\end{flalign} 
Note also that in the special case when $\left\{ \mathbf{v}_{k} \right\}$ is a white noise, i.e., when $F_{i} = G_{j} = 0$ in \eqref{ARMA}, \eqref{Cdesign2} reduces to
\begin{flalign}
C = I_{n \times n}.
\end{flalign}

\begin{figure}
	\begin{center}
		\vspace{-3mm}
		\includegraphics [width=0.5\textwidth]{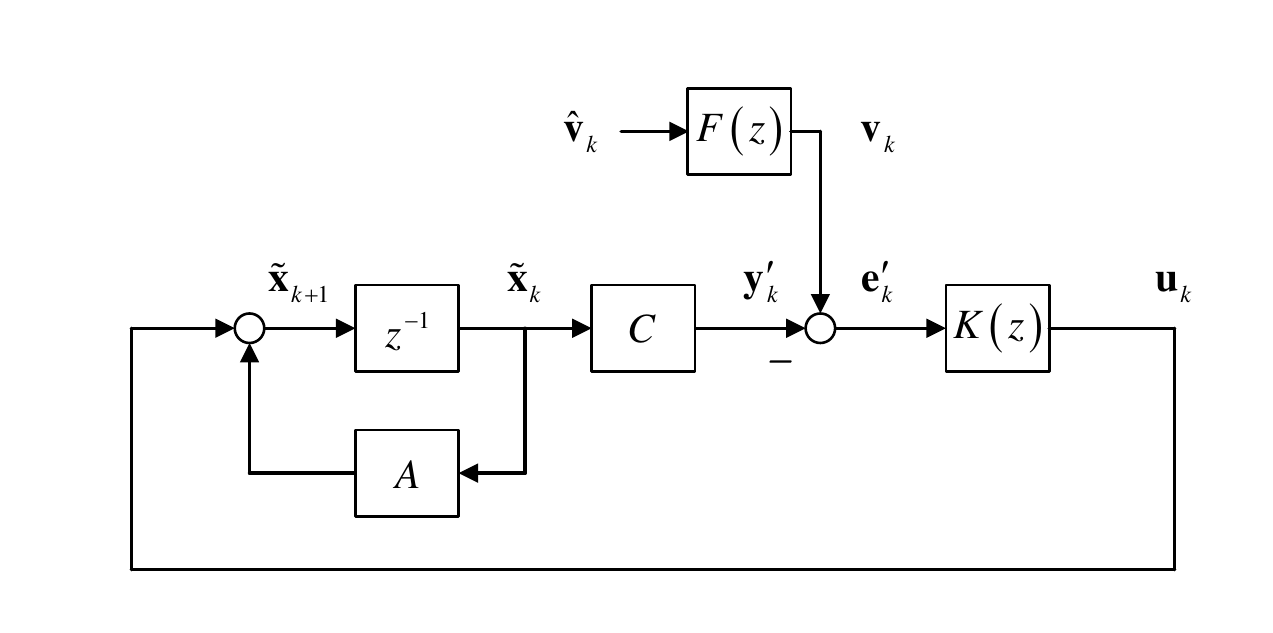}
		\vspace{-6mm}
		\caption{The steady-state integrated Kalman filter for colored noises: Equivalent form.}
		\label{kalman4}
	\end{center}
	\vspace{-3mm}
\end{figure}

On the other hand, we may obtain an equivalent form of the system in Fig.~\ref{kalman3} as given by \eqref{integrate}.

\begin{proposition} \label{observation3}
	The system in Fig.~\ref{kalman3} is equivalent to that in Fig.~\ref{kalman4}, where $K \left( z \right)$ is dynamic and is given by
	\begin{flalign}
	K \left( z \right)
	&= F^{-1} \left( z \right) \widehat{K} \nonumber \\
	&=  \left( I - \sum_{i=1}^{p} F_{i} z^{-i} \right) \left( I + \sum_{j=1}^{q} G_{j} z^{-j} \right)^{-1} \widehat{K}. 
	\end{flalign}
	Herein, $\widehat{K}$ is given by \eqref{gain2}.
	More specifically, the system in Fig.~\ref{kalman4} is given by 
	\begin{flalign} \label{integrate6}
	\left\{ \begin{array}{rcl}
	\widetilde{\mathbf{x}}_{k+1}&=&A \widetilde{\mathbf{x}}_{k} +\mathbf{u}_k, \\
	\mathbf{y}'_{k}&= & C \widetilde{\mathbf{x}}_{k}, \\
	\mathbf{e}'_{k}&=& - \mathbf{y}'_k + \mathbf{v}_k, \\
	\mathbf{u}_{k}&=& \widehat{K} \left( \mathbf{e}'_{k} - \sum_{i=1}^{p} F_{i} \mathbf{e}'_{k-i} \right) - \sum_{j=1}^{q} G_{j} \mathbf{u}_{k-j},
	\end{array}
	\right.
	\end{flalign}
	which is stable as a closed-loop system.
\end{proposition}

\begin{IEEEproof}
	Note first that the system of Fig.~\ref{kalman3} is equivalent to the one of Fig.~\ref{kalman5}, since $\widehat{K}  = F \left( z \right) K \left( z \right)$.
	\begin{figure}
		\begin{center}
			\vspace{-3mm}
			\includegraphics [width=0.5\textwidth]{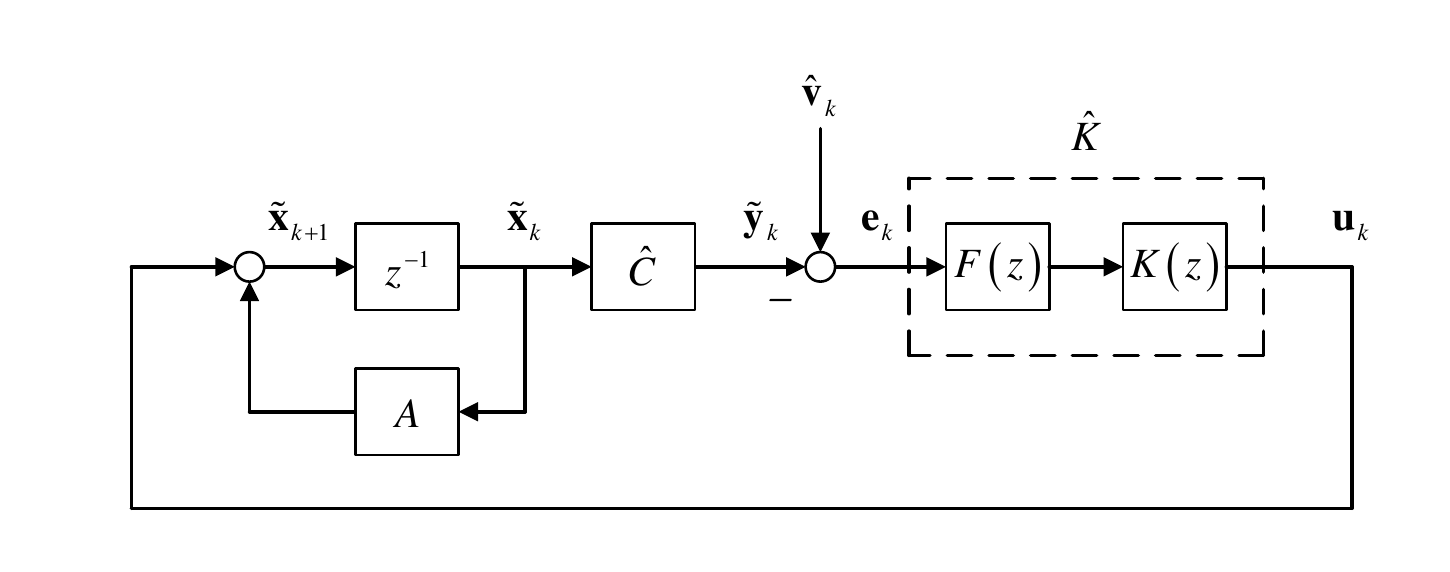}
			\vspace{-6mm}
			\caption{The steady-state integrated Kalman filter for colored noises: Equivalent form~2.}
			\label{kalman5}
		\end{center}
		\vspace{-3mm}
	\end{figure}
	In addition, it is known from the proof of Proposition~\ref{anderson1} that
	\begin{flalign}
	\widetilde{\mathbf{y}}_{k}
	&= \widehat{C} \widetilde{\mathbf{x}}_{k} =  \left( C - \sum_{i=1}^{\infty} H_{i} C A^{-i} \right) \widetilde{\mathbf{x}}_{k}. \nonumber
	\end{flalign}
	As such, since 
	\begin{flalign}
	\widetilde{\mathbf{x}}_{k} = \overline{\mathbf{x}}_{k} - \mathbf{x}_{k} = A \left( \overline{\mathbf{x}}_{k-1} - \mathbf{x}_{k-1}\right) = A \widetilde{\mathbf{x}}_{k-1}, \nonumber
	\end{flalign}
	we have
	\begin{flalign}
	&\left( C - \sum_{i=1}^{\infty} H_{i} C A^{-i} \right) \widetilde{\mathbf{x}}_{k}
	= \left( C - \sum_{i=1}^{\infty} H_{i} C z^{-i} \right) \widetilde{\mathbf{x}}_{k} \nonumber \\
	&
	~~~~ = \left( I - \sum_{i=1}^{p} F_{i} z^{-i} \right) \left( I + \sum_{j=1}^{q} G_{j} z^{-j} \right)^{-1} C \widetilde{\mathbf{x}}_{k} \nonumber \\
	&
	~~~~ = \left( I - \sum_{i=1}^{\infty} H_{i} z^{-i} \right) C \widetilde{\mathbf{x}}_{k}
	= F^{-1} \left( z \right) C  \widetilde{\mathbf{x}}_{k}. \nonumber
	\end{flalign}
	Consequently, the system of Fig.~\ref{kalman5} is equivalent to that of Fig.~\ref{kalman6}.
	\begin{figure}
		\begin{center}
			\vspace{-3mm}
			\includegraphics [width=0.5\textwidth]{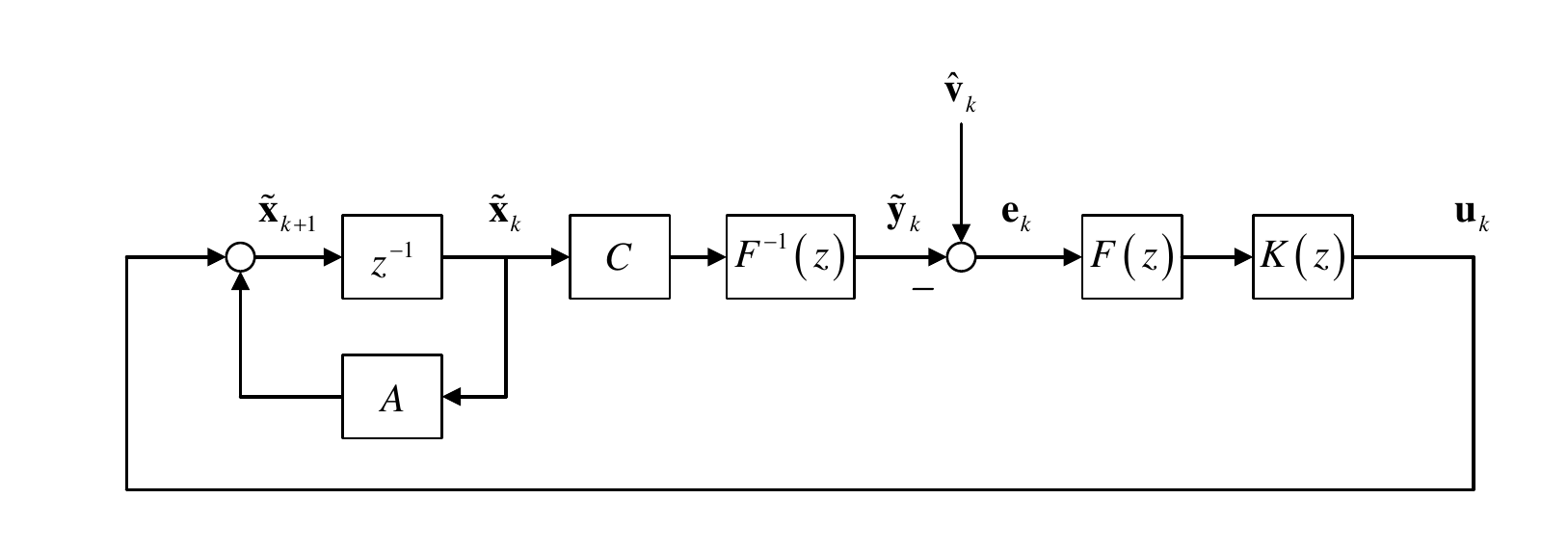}
			\vspace{-6mm}
			\caption{The steady-state integrated Kalman filter for colored noises: Equivalent form~3.}
			\label{kalman6}
		\end{center}
		\vspace{-3mm}
	\end{figure}
	Moreover, since all the sub-systems are linear, the system of Fig.~\ref{kalman6} is equivalent to that of Fig.~\ref{kalman7}, which in turn equals to the one of Fig.~\ref{kalman4}; note that herein $F \left( z \right)$ is stable and minimum-phase, and thus there will be no issues caused by cancellations of unstable poles and nonminimum-phase zeros.
	\begin{figure}
		\begin{center}
			\vspace{-3mm}
			\includegraphics [width=0.5\textwidth]{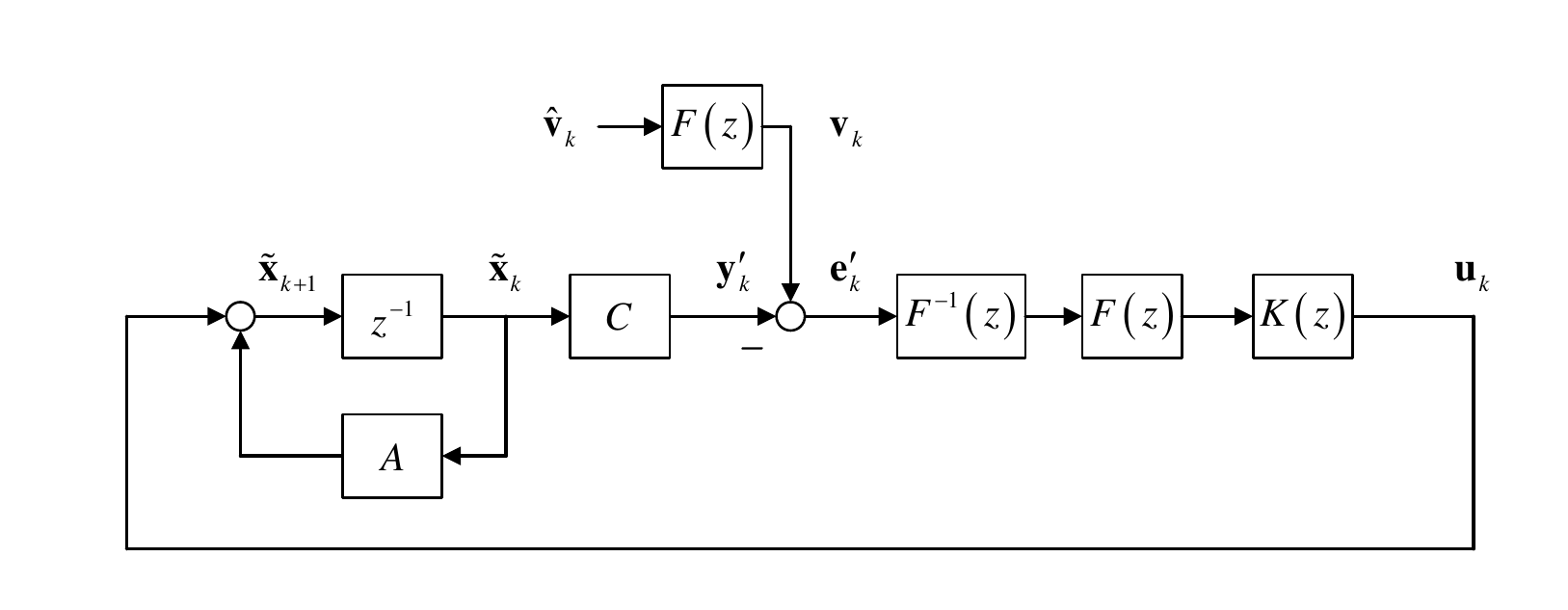}
			\vspace{-6mm}
			\caption{The steady-state integrated Kalman filter for colored noises: Equivalent form~4.}
			\label{kalman7}
		\end{center}
		\vspace{-3mm}
	\end{figure}
	Meanwhile, the closed-loop stability of the system given in \eqref{integrate6} and Fig.~\ref{kalman4} is the same as that of the system given by \eqref{integrate} and Fig.~\ref{kalman3}, since they are essentially the same feedback system.
\end{IEEEproof}


\begin{figure}
	\begin{center}
		\vspace{-3mm}
		\includegraphics [width=0.5\textwidth]{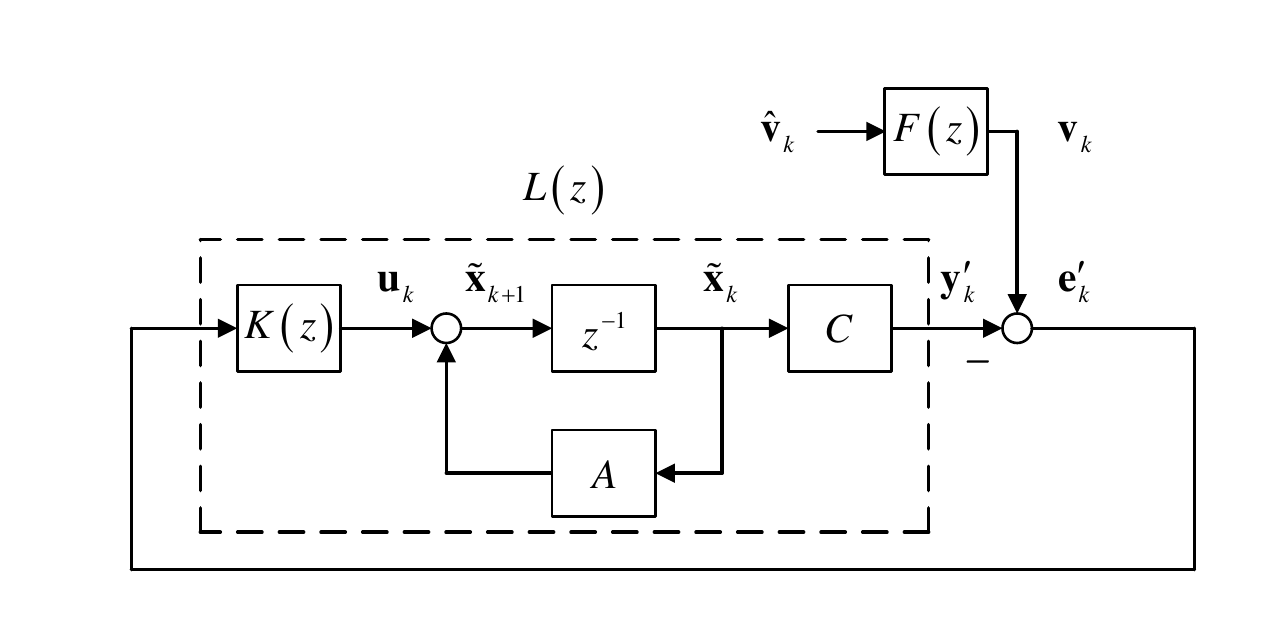}
		\vspace{-6mm}
		\caption{The steady-state integrated Kalman filter for colored noises: Equivalent form~5.}
		\label{kalman10}
	\end{center}
	\vspace{-3mm}
\end{figure}

As a matter of fact, in the system of Fig.~\ref{kalman4}, or equivalently, in the system of Fig.~\ref{kalman10}, we may view 
\begin{flalign} \label{notations}
\mathbf{e}'_{k} = - \mathbf{y}'_{k} + \mathbf{v}_{k}
\end{flalign}
as a feedback channel \cite{kim2010feedback, ardestanizadeh2012control} with additive colored Gaussian noise $\left\{ \mathbf{v}_{k} \right\}$, whereas $\left\{ - \mathbf{y}'_{k} \right\}$ is the channel input while $\left\{ \mathbf{e}'_{k} \right\}$ is the channel output. Note that in Fig.~\ref{kalman10}, 
\begin{flalign} \label{feedbackcoding}
L \left( z \right) = C \left( z I - A \right)^{-1} K \left( z \right)
\end{flalign}
may be viewed as a particular class of feedback coding.
On the other hand, with the notations in \eqref{notations}, the feedback capacity is given by (cf. the definition in \eqref{fcdef})
\begin{flalign} \label{notations2}
C_{\text{f}} = \sup_{ \lim_{k\to \infty} \frac{1}{k+1} \sum_{i=0}^{k} \tr \mathbb{E} \left[  \left( - \mathbf{y}'_{i} \right)  \left( - \mathbf{y}'_{i} \right)^{\mathrm{T}} \right] \leq \overline{P}} \left[ h_{\infty} \left( \mathbf{e}' \right) -  h_{\infty} \left(  \mathbf{v} \right) \right].
\end{flalign}
As such, if $A$ and $C$ are designed specifically as in Theorem~\ref{t1}, then \eqref{feedbackcoding} naturally provides a class of sub-optimal feedback coding scheme, by which the corresponding $h_{\infty} \left( \mathbf{e}' \right) -  h_{\infty} \left(  \mathbf{v} \right)$ that can be achieved is thus a lower bound of \eqref{notations2}.

In this view, the following lower bound of feedback capacity can be obtained.

\begin{theorem} \label{fc4}
	Suppose that in \eqref{ARMA}, $\widehat{V} = U_{\widehat{\mathbf{v}}} \Lambda_{\widehat{\mathbf{v}}} U^{\mathrm{T}}_{\widehat{\mathbf{v}}}$, where 
	\begin{flalign} 
	\Lambda_{\widehat{\mathbf{v}}} = \mathrm{diag} \left( \widehat{V}_1, \ldots, \widehat{V}_{n}\right),
	\end{flalign}
	and $U_{\widehat{\mathbf{v}}}$ is an orthogonal matrix.
	Then, a lower bound of the feedback capacity with power constraint $\overline{P}$ is given by 
	\begin{flalign} \label{max}
	\max_{P_1 \geq 0, \ldots, P_n \geq 0}
	\sum_{\ell=1}^{n} \frac{1}{2} \log \left( 1 + \frac{ P_{\ell} }{ \widehat{V}_{\ell} } \right),
	\end{flalign}
	where $P_1, \ldots, P_n$ satisfy
	\begin{flalign} \label{max2}
	\tr \left( C U_{\widehat{\mathbf{v}}} \left[
	\begin{array}{cccc}
	P_1 & \cdots & 0\\
	\vdots & \ddots & \vdots\\
	0 & \cdots & P_n\\
	\end{array} \right] U^{\mathrm{T}}_{\widehat{\mathbf{v}}} C^{\mathrm{T}} \right) = \overline{P}.
	\end{flalign}
	Herein, 
	\begin{flalign} \label{fcoding1}
	&C = \mathrm{vec}^{-1}_{n \times n} \Bigg[ \left( U_{\widehat{\mathbf{v}}}^{-\mathrm{T}} \otimes I_{n \times n} \right) \left( I_{n^2 \times n^2} + \sum_{j=1}^{q} \Lambda^{-j} \otimes G_{j} \right) \nonumber \\
	& \times \left( I_{n^2 \times n^2} - \sum_{i=1}^{p} \Lambda^{-i} \otimes F_{i} \right)^{-1} \left(U_{\widehat{\mathbf{v}}}^{\mathrm{T}} \otimes I_{n \times n} \right) \mathrm{vec} \left( I_{n \times n} \right) \Bigg]
	,
	\end{flalign} 
	where
	\begin{flalign} \label{fcoding2}
	\Lambda
	= \pm  \left[
	\begin{array}{cccc}
	\sqrt{1 + \frac{P_1}{\widehat{V}_1}} & \cdots & 0\\
	\vdots & \ddots & \vdots\\
	0 & \cdots & \sqrt{1 + \frac{P_n}{\widehat{V}_n}}\\
	\end{array} \right].
	\end{flalign}
\end{theorem}

\begin{IEEEproof}
	To start with, suppose that $A$ and $C$ are specifically designed as in Theorem~\ref{t1}. In this case, it is known from Theorem~\ref{observation3} that the system in \eqref{integrate6} is stable.
	Note then that \eqref{integrate6} implies
	\begin{flalign}
	-Y' \left( z \right) 
	&= - C \left( z I - A \right)^{-1} K \left( z \right) \nonumber \\
	&~~~~ \times \left[ I + C \left( z I - A \right)^{-1} K \left( z \right) \right]^{-1} V \left( z \right), \nonumber
	\end{flalign}
	and thus
	\begin{flalign} \label{closed}
	- C \left( z I - A \right)^{-1} K \left( z \right) \left[ I + C \left( z I - A \right)^{-1} K \left( z \right) \right]^{-1}
	\end{flalign}	
	is stable. Accordingly, since $\left\{ \mathbf{v}_{k} \right\}$ is stationary Gaussian, $\left\{ - \mathbf{y}'_{k} \right\}$ is also stationary Gaussian. 
	On the other hand, it holds that
	\begin{flalign}
	E' \left( z \right) 
	= \left[ I + C \left( z I - A \right)^{-1} K \left( z \right) \right]^{-1} V \left( z \right), \nonumber
	\end{flalign}
	and as a consequence (cf. discussions in \cite{Eli:04, ardestanizadeh2012control}),
	\begin{flalign}
	&h_{\infty} \left( \mathbf{e}' \right) - h_{\infty} \left( \mathbf{v} \right) \nonumber \\
	&= \frac{1}{2 \pi} \int_{- \pi}^{\pi} \log \det \left| \left[ I + C \left( \mathrm{e}^{\mathrm{j} \omega} I - A \right)^{-1} K \left( \mathrm{e}^{\mathrm{j} \omega} \right) \right]^{-1} \right| \mathrm{d} \omega \nonumber \\
	&= \frac{1}{2 \pi} \int_{- \pi}^{\pi} \log \det \left| \left[ I + C \left( \mathrm{e}^{\mathrm{j} \omega} I - A \right)^{-1} F^{-1} \left( \mathrm{e}^{\mathrm{j} \omega} \right) \widehat{K} \right]^{-1} \right| \mathrm{d} \omega \nonumber \\
	&= \sum_{\ell=1}^{n} \log \left| \lambda_{\ell} \right| = \sum_{\ell=1}^{n} \log \sqrt{ 1 + \frac{P_{\ell}}{\widehat{V}_{\ell}} } = \sum_{\ell=1}^{n} \frac{1}{2} \log \left( 1 + \frac{P_{\ell}}{\widehat{V}_{\ell}} \right)
	, \nonumber
	\end{flalign}
	where the first equality may be referred to \cite{Pap:02, fang2017towards} while the third equality follows as a result of the Bode integral or Jensen's formula \cite{seron2012fundamental, fang2017towards}. Note that herein we have used the fact that $F^{-1} \left( z \right)$ is stable and minimum-phase,  $\left( A, C \right)$ is detectable (thus the set of unstable poles of $C \left( z I - A \right)^{-1} K \left( z \right)$
	is exactly the same as the set of eigenvalues of $A$ with magnitude greater than or equal to $1$; see, e.g., discussions in \cite{FangACC18}), and 
	\begin{flalign}
	\left[ I + C \left( z I - A \right)^{-1} K \left( z \right) \right]^{-1} \nonumber
	\end{flalign}
	is stable.
	Consequently, according to the definition of feedback capacity given in \eqref{notations2}, it holds that
	\begin{flalign}
	C_{\mathrm{f}}
	\geq h_{\infty} \left( \mathbf{e}' \right) - h_{\infty} \left( \mathbf{v} \right)
	= \sum_{\ell=1}^{n} \frac{1}{2} \log \left( 1 + \frac{P_{\ell}}{\widehat{V}_{\ell}} \right)
	, \nonumber
	\end{flalign}
	when the corresponding 
	\begin{flalign}
	&\lim_{k\to \infty} \frac{1}{k+1} \sum_{i=0}^{k} \tr \mathbb{E} \left[  \left( - \mathbf{y}'_{i} \right)  \left( - \mathbf{y}'_{i} \right)^{\mathrm{T}} \right] \nonumber \\
	&~~~~ = \tr \mathbb{E} \left[ \left( - \mathbf{y}'_{k} \right) \left( - \mathbf{y}'_{k} \right)^{\mathrm{T}} \right] 
	= \tr \mathbb{E}  \left[ \left( \mathbf{y}'_{k} \right) \left( \mathbf{y}'_{k} \right)^{\mathrm{T}} \right] \nonumber \\
	&~~~~ = \tr \mathbb{E}  \left[ \left( C \widetilde{\mathbf{x}}_{k} \right) \left( C \widetilde{\mathbf{x}}_{k} \right)^{\mathrm{T}} \right] 
	= \tr \mathbb{E}  \left[ C \widetilde{\mathbf{x}}_{k} \widetilde{\mathbf{x}}_{k} ^{\mathrm{T}} C^{\mathrm{T}} \right] 
	\nonumber \\
	&~~~~ = \tr \left\{ C \mathbb{E}  \left[ \left( \overline{\mathbf{x}}_{k} - \mathbf{x}_{k} \right) \left( \overline{\mathbf{x}}_{k} - \mathbf{x}_{k} \right) ^{\mathrm{T}} \right] C^{\mathrm{T}} \right\}
	= \tr \left( C P C^{\mathrm{T}} \right) \nonumber
	\end{flalign} 
	is less than the power constraint $\overline{P}$, i.e., when (see Theorem~\ref{c1})
	\begin{flalign}
	\tr \left( C P C^{\mathrm{T}} \right)
	= \tr \left( C U_{\widehat{\mathbf{v}}} \left[
	\begin{array}{cccc}
	P_1 & \cdots & 0\\
	\vdots & \ddots & \vdots\\
	0 & \cdots & P_n\\
	\end{array} \right] U^{\mathrm{T}}_{\widehat{\mathbf{v}}} C^{\mathrm{T}} \right) \leq \overline{P}. \nonumber
	\end{flalign}
	Note that herein we have used the fact that $\left\{ - \mathbf{y}'_{k} \right\}$ is stationary.
	In particular, we may pick the 	
	allocation
	$P_{1}, \ldots, P_{n} $ that maximizes 
	\begin{flalign}
	\sum_{\ell=1}^{n} \frac{1}{2} \log \left( 1 + \frac{P_{\ell}}{\widehat{V}_{\ell}} \right)
	\nonumber
	\end{flalign}
	while satisfying \begin{flalign}
	\tr \left( C U_{\widehat{\mathbf{v}}} \left[
	\begin{array}{cccc}
	P_1 & \cdots & 0\\
	\vdots & \ddots & \vdots\\
	0 & \cdots & P_n\\
	\end{array} \right] U^{\mathrm{T}}_{\widehat{\mathbf{v}}} C^{\mathrm{T}} \right) = \overline{P}. \nonumber
	\end{flalign} 
	This completes the proof.
\end{IEEEproof}


Note that the solution to \eqref{max} is essentially a power allocation policy with feedback. Note also that he lower bound in Theorem~\ref{fc4} is equal to
\begin{flalign} 
\max_{a_1 \geq 1, \ldots, a_n \geq 1}
\sum_{\ell=1}^{n} \log a_{\ell},
\end{flalign}
where $a_1, \ldots, a_n$ satisfy
\begin{flalign} 
\tr \left( C U_{\widehat{\mathbf{v}}} \left[
\begin{array}{cccc}
a_1^2 - 1 & \cdots & 0\\
\vdots & \ddots & \vdots\\
0 & \cdots & a_n^2 - 1\\
\end{array} \right] U^{\mathrm{T}}_{\widehat{\mathbf{v}}} C^{\mathrm{T}} \right) = \overline{P}.
\end{flalign}
Herein, $C$ is given by \eqref{fcoding1},
where
\begin{flalign}
\Lambda
= \pm \left[
\begin{array}{cccc}
a_1 & \cdots & 0\\
\vdots & \ddots & \vdots\\
0 & \cdots & a_n\\
\end{array} \right].
\end{flalign}

We now consider the case of independent parallel channels.

\begin{corollary} \label{c1}
	Suppose that in \eqref{ARMA}, 
	\begin{flalign} 
	\widehat{V} = \mathrm{diag} \left( \widehat{V}_1, \ldots, \widehat{V}_{n}\right),
	\end{flalign}
	and 
	\begin{flalign} 
	F_i = \mathrm{diag} \left( f_{i1}, \ldots, f_{in}\right),~i=1,\ldots,p,
	\end{flalign}
	while 
	\begin{flalign} 
	G_j = \mathrm{diag} \left( g_{j1}, \ldots, g_{jn}\right),~j=1,\ldots,q,
	\end{flalign}
	which essentially model a parallel of independent ARMA noises. 
	Then, a lower bound of the feedback capacity with power constraint $\overline{P}$ is given by 
	\begin{flalign}
	\max_{P_1 \geq 0, \ldots, P_n \geq 0}
	\sum_{\ell=1}^{n} \frac{1}{2} \log \left( 1 + \frac{ P_{\ell} }{ \widehat{V}_{\ell} } \right),
	\end{flalign}
	where $P_1, \ldots, P_n$ satisfy
	\begin{flalign} \left[
	\sum_{\ell=1}^{n} \left( \frac{1 + \sum_{j=1}^{q} g_{j\ell} a_{\ell}^{-j} }{ 1 - \sum_{i=1}^{p} f_{i\ell} a_{\ell}^{-i} } \right)^2 P_{\ell} \right] = \overline{P},
	\end{flalign}
	or 
	\begin{flalign} 
	\sum_{\ell=1}^{n} \left\{ \left[ \frac{1 + \sum_{j=1}^{q} g_{j\ell} \left(- a_{\ell} \right)^{-j} }{ 1 - \sum_{i=1}^{p} f_{i\ell} \left(- a_{\ell} \right)^{-i} } \right]^2 P_{\ell} \right\}
	= \overline{P} .
	\end{flalign}
	Herein,
	\begin{flalign} \label{aP}
	a_{\ell}
	= \sqrt{1 + \frac{P_{\ell}}{\widehat{V}_{\ell}}}, \ell=1,\ldots,n.
	\end{flalign}
\end{corollary}

\begin{IEEEproof}
	Note first that in this case, $U_{\widehat{\mathbf{v}}} = I$ and hence
	\begin{flalign}
	&C = \mathrm{vec}^{-1}_{n \times n} \Bigg[ \left( I_{n \times n}^{-\mathrm{T}} \otimes I_{n \times n} \right) \left( I_{n^2 \times n^2} + \sum_{j=1}^{q} \Lambda^{-j} \otimes G_{j} \right) \nonumber \\
	& \times \left( I_{n^2 \times n^2} - \sum_{i=1}^{p} \Lambda^{-i} \otimes F_{i} \right)^{-1} \left(I_{n \times n}^{\mathrm{T}} \otimes I_{n \times n} \right) \mathrm{vec} \left( I_{n \times n} \right) \Bigg] \nonumber \\
	&~~~~ = \mathrm{vec}^{-1}_{n \times n} \Bigg[ \left( I_{n^2 \times n^2} + \sum_{j=1}^{q} \Lambda^{-j} \otimes G_{j} \right) \nonumber \\
	&~~~~~~~~ \times \left( I_{n^2 \times n^2} - \sum_{i=1}^{p} \Lambda^{-i} \otimes F_{i} \right)^{-1}  \mathrm{vec} \left( I_{n \times n} \right) \Bigg] \nonumber
	,
	\end{flalign} 
	while
	\begin{flalign}
	&\tr \left( C U_{\widehat{\mathbf{v}}} \left[
	\begin{array}{cccc}
	P_1 & \cdots & 0\\
	\vdots & \ddots & \vdots\\
	0 & \cdots & P_n\\
	\end{array} \right] U^{\mathrm{T}}_{\widehat{\mathbf{v}}} C^{\mathrm{T}} \right) \nonumber \\
	&~~~~ = \tr \left( C  \left[
	\begin{array}{cccc}
	P_1 & \cdots & 0\\
	\vdots & \ddots & \vdots\\
	0 & \cdots & P_n\\
	\end{array} \right] C^{\mathrm{T}} \right). \nonumber 
	\end{flalign}
	
	As such, if 
	\begin{flalign} 
	\Lambda = \mathrm{diag} \left( a_{1}, \ldots, a_{n}\right), \nonumber
	\end{flalign}	
	where $a_{1}, \ldots, a_{n}$ are given by \eqref{aP}, then similarly to the procedures in the proof of Proposition~\ref{anderson1}, it can be obtained that
	\begin{flalign}
	&\left( I_{n^2 \times n^2} + \sum_{j=1}^{q} \Lambda^{-j} \otimes G_{j} \right) \left( I_{n^2 \times n^2} - \sum_{i=1}^{p} \Lambda^{-i} \otimes F_{i} \right)^{-1} \nonumber \\
	&= \left[
	\begin{array}{ccc}
	I_{n \times n} + \sum_{j=1}^{q} a_{1}^{-j} G_{j} & \cdots & 0\\
	\vdots & \ddots & \vdots\\
	0 & \cdots & I_{n \times n} + \sum_{j=1}^{q} a_{n}^{-j} G_{j} \\
	\end{array} \right] \nonumber \\
	&\times \left[
	\begin{array}{ccc}
	I_{n \times n} - \sum_{i=1}^{p} a_{1}^{-i} F_{i} & \cdots & 0\\
	\vdots & \ddots & \vdots\\
	0 & \cdots & I_{n \times n} - \sum_{i=1}^{p} a_{n}^{-i} F_{i} \\
	\end{array} \right]^{-1}. \nonumber
	\end{flalign}
	Meanwhile, in this case it holds for $\ell = 1, \ldots, n,$ that
	\begin{flalign}
	&I_{n \times n} + \sum_{j=1}^{q} a_{\ell}^{-j} G_{j} \nonumber \\
	&~~~~ = \left[
	\begin{array}{ccc}
	1 + \sum_{j=1}^{q} a_{\ell}^{-j} g_{j1} & \cdots & 0\\
	\vdots & \ddots & \vdots\\
	0 & \cdots & 1 + \sum_{j=1}^{q} a_{\ell}^{-j} g_{jn} \\
	\end{array} \right], \nonumber
	\end{flalign}
	and
	\begin{flalign}
	&\left(  I_{n \times n} - \sum_{i=1}^{p} a_{\ell}^{-i} F_{i} \right)^{-1} \nonumber \\
	&~~~~ = \left[
	\begin{array}{ccc}
	1 - \sum_{i=1}^{p} a_{\ell}^{-i} f_{i1} & \cdots & 0\\
	\vdots & \ddots & \vdots\\
	0 & \cdots & 1 - \sum_{i=1}^{p} a_{\ell}^{-i} f_{in} \\
	\end{array} \right]^{-1}. \nonumber
	\end{flalign}
	Thus,
	\begin{flalign}
	C & = \mathrm{vec}^{-1}_{n \times n} \Bigg[ \left( I_{n^2 \times n^2} + \sum_{j=1}^{q} \Lambda^{-j} \otimes G_{j} \right) \nonumber \\
	&~~~~ \times \left( I_{n^2 \times n^2} - \sum_{i=1}^{p} \Lambda^{-i} \otimes F_{i} \right)^{-1}  \mathrm{vec} \left( I_{n \times n} \right) \Bigg] \nonumber \\
	&= \left[
	\begin{array}{ccc}
	\frac{1 + \sum_{j=1}^{q} g_{j1} a_{1}^{-j} }{ 1 - \sum_{i=1}^{p} f_{i1} a_{1}^{-i} } & \cdots & 0\\
	\vdots & \ddots & \vdots\\
	0 & \cdots & \frac{1 + \sum_{j=1}^{q} g_{jn} a_{n}^{-j} }{ 1 - \sum_{i=1}^{p} f_{in} a_{n}^{-i} } \\
	\end{array} \right], \nonumber
	\end{flalign} 
	and the power constraint becomes
	\begin{flalign}
	&\tr \left( C  \left[
	\begin{array}{cccc}
	P_1 & \cdots & 0\\
	\vdots & \ddots & \vdots\\
	0 & \cdots & P_n\\
	\end{array} \right] C^{\mathrm{T}} \right) \nonumber \\
	&~~~~ = 
	\sum_{\ell=1}^{n} \left[ \left( \frac{1 + \sum_{j=1}^{q} g_{j\ell} a_{\ell}^{-j} }{ 1 - \sum_{i=1}^{p} f_{i\ell} a_{\ell}^{-i} } \right)^2 P_{\ell} \right] = \overline{P}
	. \nonumber 
	\end{flalign}
	
	Similarly, if 
	\begin{flalign} 
	\Lambda = - \mathrm{diag} \left( a_{1}, \ldots, a_{n}\right), \nonumber
	\end{flalign}
	then it can be obtained that	
	\begin{flalign}
	C = \left[
	\begin{array}{ccc}
	\frac{1 + \sum_{j=1}^{q} g_{j1} \left( - a_{1} \right)^{-j} }{ 1 - \sum_{i=1}^{p} f_{i1} \left( - a_{1} \right)^{-i} } & \cdots & 0\\
	\vdots & \ddots & \vdots\\
	0 & \cdots & \frac{1 + \sum_{j=1}^{q} g_{jn} \left( - a_{n} \right)^{-j} }{ 1 - \sum_{i=1}^{p} f_{in} \left( - a_{n} \right)^{-i} } \\
	\end{array} \right], \nonumber
	\end{flalign} 
	and the power constraint becomes
	\begin{flalign}
	\sum_{\ell=1}^{n} \left\{ \left[ \frac{1 + \sum_{j=1}^{q} g_{j} \left( -a \right)^{-j} }{ 1 - \sum_{i=1}^{p} f_{i} \left( -a \right)^{-i} } \right]^2 P_{\ell} \right\} = \overline{P}
	. \nonumber 
	\end{flalign}
	This completes the proof.
\end{IEEEproof}

Equivalently, the lower bound in Corollary~\ref{c1} can be rewritten as
\begin{flalign} 
\max_{a_1 \geq 1, \ldots, a_n \geq 1}
\sum_{\ell=1}^{n} \log a_{\ell},
\end{flalign}
where $a_1, \ldots, a_n$ satisfy
\begin{flalign} 
\sum_{\ell=1}^{n} \left[ \left( \frac{1 + \sum_{j=1}^{q} g_{j\ell} a_{\ell}^{-j} }{ 1 - \sum_{i=1}^{p} f_{i\ell} a_{\ell}^{-i} } \right)^2 \left( a_{\ell}^2 - 1 \right) \widehat{V}_{\ell} \right] = \overline{P},
\end{flalign}
or 
\begin{flalign} 
\sum_{\ell=1}^{n} \left\{ \left[ \frac{1 + \sum_{j=1}^{q} g_{j\ell} \left(- a_{\ell} \right)^{-j} }{ 1 - \sum_{i=1}^{p} f_{i\ell} \left(- a_{\ell} \right)^{-i} } \right]^2 \left( a_{\ell}^2 - 1 \right) \widehat{V}_{\ell} \right\}
= \overline{P}.
\end{flalign}

We next consider some special cases in which Theorem~\ref{fc4} (or Corollary~\ref{c1}) can be characterized more explicitly, including a parallel of AWGN channels (Example~\ref{e1}) and a single ACGN channel (Example~~\ref{e2}).

\begin{example} \label{e1}
	In the special case when $\left\{ \mathbf{v}_{k} \right\}$ is a white Gaussian noise with covariance $\widehat{V}$, i.e., when $F_{i} = G_{j} = 0$ in \eqref{ARMA}, 
	a lower bound of the feedback capacity with power constraint $\overline{P}$ is given by 
	\begin{flalign} \label{water}
	\max_{P_1 \geq 0, \ldots, P_n \geq 0}
	\sum_{\ell=1}^{n} \frac{1}{2} \log \left( 1 + \frac{ P_{\ell} }{ \widehat{V}_{\ell} } \right),
	\end{flalign}
	where $P_1, \ldots, P_n$ satisfy
	\begin{flalign} 
	\tr \left( U_{\widehat{\mathbf{v}}} \left[
	\begin{array}{cccc}
	P_1 & \cdots & 0\\
	\vdots & \ddots & \vdots\\
	0 & \cdots & P_n\\
	\end{array} \right] U^{\mathrm{T}}_{\widehat{\mathbf{v}}} \right)
	& = \tr \left( \left[
	\begin{array}{cccc}
	P_1 & \cdots & 0\\
	\vdots & \ddots & \vdots\\
	0 & \cdots & P_n\\
	\end{array} \right] \right) \nonumber \\
	&= \sum_{\ell=1}^{n} P_{\ell} = \overline{P}.
	\end{flalign}
	As a matter of fact, the lower bound is tight in this case \cite{Cov:06} and the optimal power allocation solution is given by the classical ``water-filling'' policy \cite{Cov:06} as
	\begin{flalign}
	P_{\ell} = \max \left\{ 0, \zeta - \widehat{V}_{\ell} \right\}, \ell=1,\ldots,n,
	\end{flalign}
	where $\zeta > 0$ satisfies
	\begin{flalign} 
	\sum_{\ell=1}^{n} P_{\ell} 
	= \sum_{\ell=1}^{n} \max \left\{ 0, \zeta - \widehat{V}_{\ell} \right\}
	= \overline{P}.
	\end{flalign}
	It is also worth mentioning that the lower bound in \eqref{water} can equivalently be rewritten as (cf. also discussions after Theorem~\ref{fc4} for the general case)
	\begin{flalign} 
	\max_{a_1 \geq 1, \ldots, a_n \geq 1}
	\sum_{\ell=1}^{n} \log a_{\ell},
	\end{flalign}
	where $a_1, \ldots, a_n$ satisfy
	\begin{flalign} 
	\sum_{\ell=1}^{n} \left[ \left( a_{\ell}^2 - 1 \right) \widehat{V}_{\ell} \right] = \overline{P}.
	\end{flalign}
	Correspondingly, the optimal ``allocation'' solution is given by
	\begin{flalign}
	a_{\ell} = \sqrt{ \max \left\{ 1, \frac{\zeta}{\widehat{V}_{\ell}} \right\} }, \ell=1,\ldots,n,
	\end{flalign}
	where $\zeta > 0$ satisfies
	\begin{flalign} 
	\sum_{\ell=1}^{n} \left[ \left( a_{\ell}^2 - 1 \right) \widehat{V}_{\ell} \right] = \sum_{\ell=1}^{n} \max \left\{ 0, \zeta - \widehat{V}_{\ell} \right\}
	= \overline{P}.
	\end{flalign}
	This provides an alternative perspective to look at the water-filling allocation, while also displaying more clearly the connections with lower bounds in other cases, e.g., that of the subsequent Example~\ref{e2}.
\end{example}

\begin{example} \label{e2}
	For another special case, consider the scalar case of $n=1$. In this case, \eqref{ARMA} reduces to
	\begin{flalign} 
	\mathbf{v}_{k} 
	&= \sum_{i=1}^{p} f_{i} \mathbf{v}_{k-i} + \widehat{\mathbf{v}}_k + \sum_{j=1}^{q} g_{j} \widehat{\mathbf{v}}_{k-j} 
	, 
	\end{flalign}
	where $\left\{ \widehat{\mathbf{v}}_k \right\}, \widehat{\mathbf{v}}_k \in \mathbb{R}$ is white Gaussian with variance $\sigma_{\widehat{\mathbf{v}}}^2 > 0$. Accordingly,  Theorem~\ref{fc4} reduces to that a lower bound of the feedback capacity with power constraint $\overline{P}$ is given by 
	\begin{flalign} \label{scalar}
	\max_{P}
	\frac{1}{2} \log \left( 1 + \frac{ P }{ \sigma_{\widehat{\mathbf{v}}}^2 } \right),
	\end{flalign}
	where $P$ satisfies
	\begin{flalign} 
	\left( \frac{1 + \sum_{j=1}^{q} g_{j} a^{-j} }{ 1 - \sum_{i=1}^{p} f_{i} a^{-i} } \right)^2 P = \overline{P},
	\end{flalign}
	or
	\begin{flalign} 
	\left[ \frac{1 + \sum_{j=1}^{q} g_{j} \left( -a \right)^{-j} }{ 1 - \sum_{i=1}^{p} f_{i} \left( -a \right)^{-i} } \right]^2 P = \overline{P}.
	\end{flalign}
	Herein,
	\begin{flalign} 
	a
	= \sqrt{1 + \frac{P}{\sigma_{\widehat{\mathbf{v}}}^2}}.
	\end{flalign}
	It may then be verified that this lower bound can equivalently be rewritten as (cf. also discussions after Theorem~\ref{fc4} or Corollary~\ref{c1})
	\begin{flalign} 
	\max_{a \geq 1}
	\log a,
	\end{flalign}
	where $a$ satisfies
	\begin{flalign} 
	\left( \frac{1 + \sum_{j=1}^{q} g_{j} a^{-j} }{ 1 - \sum_{i=1}^{p} f_{i} a^{-i} } \right)^2 \left( a^2 - 1 \right) \sigma_{\widehat{\mathbf{v}}}^2 = \overline{P},
	\end{flalign}
	or
	\begin{flalign} 
	\left[ \frac{1 + \sum_{j=1}^{q} g_{j} \left( -a \right)^{-j} }{ 1 - \sum_{i=1}^{p} f_{i} \left( -a \right)^{-i} } \right]^2 \left( a^2 - 1 \right) \sigma_{\widehat{\mathbf{v}}}^2 = \overline{P}.
	\end{flalign}
	In fact, \eqref{scalar} coincides with the lower bound in \cite{fang2020connection} given as
	\begin{flalign} 
	\max_{a \in \mathbb{R}}
	\log \left| a \right|,
	\end{flalign}
	where $a$ satisfies
	\begin{flalign} 
	\left( \frac{1 + \sum_{j=1}^{q} g_{j} a^{-j} }{ 1 - \sum_{i=1}^{p} f_{i} a^{-i} } \right)^2 \left( a^2 - 1 \right) \sigma_{\widehat{\mathbf{v}}}^2 = \overline{P},
	\end{flalign}
	whereas this in turn reduces to the results in, e.g., \cite{Eli:04} (see detailed discussions in \cite{fang2020connection}, which also relates to the formulae in, e.g., \cite{kim2010feedback}).
\end{example}

\begin{figure}
	\begin{center}
		\vspace{-3mm}
		\includegraphics [width=0.3\textwidth]{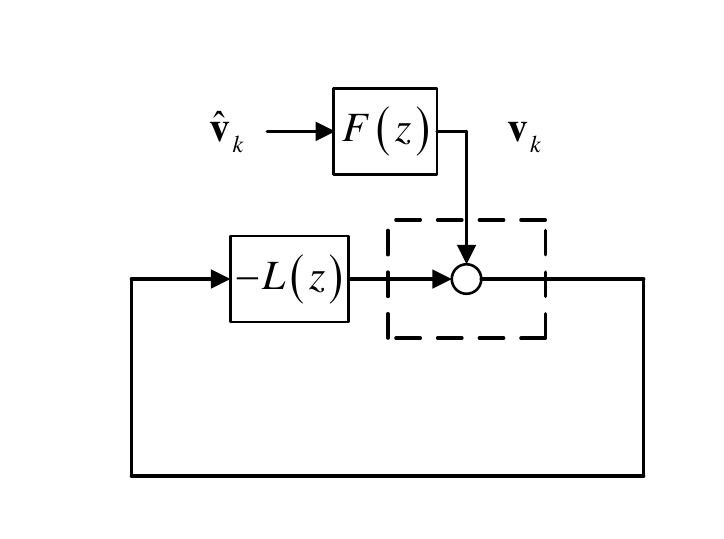}
		\vspace{-3mm}
		\caption{The steady-state integrated Kalman filter as a feedback coding scheme}
		\label{kalman11}
	\end{center}
	\vspace{-3mm}
\end{figure}

Note that \eqref{integrate6} and Fig.~\ref{kalman10} essentially provide a recursive coding scheme to achieve the lower bound in Theorem~\ref{fc4}. This is more clearly seen in Fig.~\ref{kalman11}, where $L \left( z \right)$ is given by \eqref{feedbackcoding}.
More specifically, the recursive coding algorithm is given as follows.

\begin{theorem}  \label{coding2}
	Suppose the optimal solution to \eqref{max} is given by $\overline{P}_1, \ldots, \overline{P}_n$.
	Then, one class of recursive coding scheme to achieve the lower bound in Theorem~\ref{fc4} is given by
	\begin{flalign} \label{coding}
	\left\{ \begin{array}{rcl}
	\widetilde{\mathbf{x}}_{k+1}&=&A \widetilde{\mathbf{x}}_{k} +\mathbf{u}_k, \\
	\mathbf{y}'_{k}&= & C \widetilde{\mathbf{x}}_{k}, \\
	\mathbf{e}'_{k}&=& - \mathbf{y}'_k + \mathbf{v}_k, \\
	\mathbf{u}_{k}&=& \widehat{K} \left( \mathbf{e}'_{k} - \sum_{i=1}^{p} F_{i} \mathbf{e}'_{k-i} \right) - \sum_{j=1}^{q} G_{j} \mathbf{u}_{k-j}.
	\end{array}
	\right.
	\end{flalign}
	Herein, $A = U_{\widehat{\mathbf{v}}} \Lambda U^{\mathrm{T}}_{\widehat{\mathbf{v}}}$ and $C$ is given by \eqref{fcoding1}, where $\Lambda$ is given by 
	\begin{flalign}
	\Lambda
	= \pm \left[
	\begin{array}{cccc}
	\sqrt{1 + \frac{\overline{P}_1}{\widehat{V}_1}} & \cdots & 0\\
	\vdots & \ddots & \vdots\\
	0 & \cdots & \sqrt{1 + \frac{\overline{P}_n}{\widehat{V}_n}}\\
	\end{array} \right].
	\end{flalign} 
\end{theorem}

In the case of parallel AWGN channels, Theorem~\ref{coding2} reduces to a recursive water-filling scheme.

\begin{example}
	In the special case when $\left\{ \mathbf{v}_{k} \right\}$ is a white noise, i.e., when $F_{i} = G_{j} = 0$ in \eqref{ARMA}, the coding scheme of \eqref{coding} reduces to 
	\begin{flalign} 
	\left\{ \begin{array}{rcl}
	\widetilde{\mathbf{x}}_{k+1}&=&A \widetilde{\mathbf{x}}_{k} +\mathbf{u}_k, \\
	\mathbf{y}'_{k}&= & C \widetilde{\mathbf{x}}_{k}, \\
	\mathbf{e}'_{k}&=& - \mathbf{y}'_k + \mathbf{v}_k, \\
	\mathbf{u}_{k}&=& \widehat{K} \mathbf{e}'_{k}.
	\end{array}
	\right.
	\end{flalign}
	Herein, $A = U_{\widehat{\mathbf{v}}} \Lambda U^{\mathrm{T}}_{\widehat{\mathbf{v}}}$ and $C = I_{n \times n}$, where $\Lambda$ is given by 
	\begin{flalign}
	\Lambda
	= \pm \left[
	\begin{array}{cccc}
	\sqrt{ \max \left\{ 1, \frac{\zeta}{\widehat{V}_{1}} \right\} }  & \cdots & 0\\
	\vdots & \ddots & \vdots\\
	0 & \cdots & \sqrt{ \max \left\{ 1, \frac{\zeta}{\widehat{V}_{n}} \right\} }\\
	\end{array} \right],
	\end{flalign} 
	and $\zeta > 0$ satisfies
	\begin{flalign} 
	\sum_{\ell=1}^{n} \max \left\{ 0, \zeta - \widehat{V}_{\ell} \right\}
	= \overline{P}.
	\end{flalign}
	Note that this is essentially a feedback (``closed-loop'') water-filling power allocation scheme, which is potentially more ``robust'' than the classical ``open-loop'' water-filling policy; cf. results in \cite{fong2017tight} for instance. We will, however, leave detailed discussions on this topic to future research.
\end{example}

\section{Conclusion}

In this paper, from the perspective of a variant of the Kalman filter, we have obtained lower bounds on the feedback capacity of parallel ACGN channels and the accompanying recursive coding schemes in terms of power allocation policies with feedback.	
Possible future research directions include investigating the tightness of the lower bounds, as well as the special cases in which more explicit solutions (cf. water-filling) to the feedback power allocation policies may be derived.

\balance
\bibliographystyle{IEEEtran}
\bibliography{references}

\end{document}